%% file: Skeletor.tex
\documentclass[twocolumn]{article}

\usepackage[T1]{fontenc}
\usepackage[utf8]{inputenc}
\usepackage{url}
\usepackage{amsmath}
\usepackage{amssymb}
\usepackage{clrscode3e}
\usepackage{algorithm}
\usepackage{url}
\usepackage{graphicx}
\newcommand{\shortcite}[1]{\cite{#1}}
\DeclareMathOperator*{\argmax}{arg\,max}

\newcommand{\R}{\mathbb{R}}
\newcommand{\skel}{\mathrm{skel}}

\newcommand{\val}[1]{\text{val}(#1)}

\begin{document}
\title{Skeletonization via Local Separators}
\author{Andreas Bærentzen and Eva Rotenberg \\
Dept. of Applied Mathematics and Computer Science, Technical University of Denmark}
\maketitle

%
%
\begin{abstract}
We propose a new algorithm for curve skeleton computation which differs from previous algorithms by being based on the notion of \emph{local separators}. The main benefits of this approach are that it is able to capture relatively fine details and that it works robustly on a range of shape representations. Specifically, our method works on shape representations that can be construed as a spatially embedded graphs. Such representations include meshes, volumetric shapes, and graphs computed from point clouds.
We describe a simple pipeline where geometric data is initially converted to a graph, optionally simplified, local separators are computed and selected, and finally a skeleton is constructed. We test our pipeline on polygonal meshes, volumetric shapes, and point clouds. Finally, we compare our results to other methods for skeletonization according to performance and quality.
\end{abstract}

\input{Introduction.tex}
\input{Algorithm.tex}
\input{Testing.tex}

%
\input{Discussion.tex}
%
\section*{Acknowledgements}
We are indebted to Rasmus Reinhold Paulsen, Patrick Møller Jensen, Tim Felle Olsen, and Niels Jeppesen for providing  several types of data and testing the method on that data. We are grateful to Ebba Dellwik for providing the LiDAR scan of a botanical tree and to Ida Bukh Villesen for her work on a method for constructing a graph from this point set.
We thank Aasa Feragen and Rasmus Reinhold Paulsen for their many useful comments and ideas on the exposition of the paper. 


\vspace{4em} 

\bibliography{references}
\end{document}

%% file: Introduction.tex
\section{Introduction}
Many of the common shape representations that we use in computer graphics are, in  essence, spatially embedded graphs. The polygonal mesh representation is an obvious example since a polygonal mesh is usually understood to be a graph embeddable in a surface of appropriate genus. Moreover, a voxel grid can also be seen as a graph, and it is easy to create a graph from a point cloud by connecting each point to nearby points. Thus, it is not surprising that graph algorithms are often useful in geometry processing, but we also rely on, and enforce, more specific constraints on the representations. For instance, since we tend to require manifoldness of surface meshes, we effectively impose strict limitations on the connectivity: if we were to connect two arbitrary, hitherto unconnected, vertices in a triangle mesh, the result would still be a spatially embedded graph, but it would not be manifold -- actually, it would not even be a triangle mesh due to the added edge. For many mesh algorithms, that would invalidate the object as input. The morale seems to be that we usually have to abide by the constraints of the geometry representation. However, there are cases where we can relax these constraints and create algorithms which operate on the broader class of spatially embedded graphs. The benefit of this would be that we might obtain algorithms applicable to a wider range of inputs.
\begin{figure}[h!]
    \centering
    \includegraphics[width=\columnwidth]{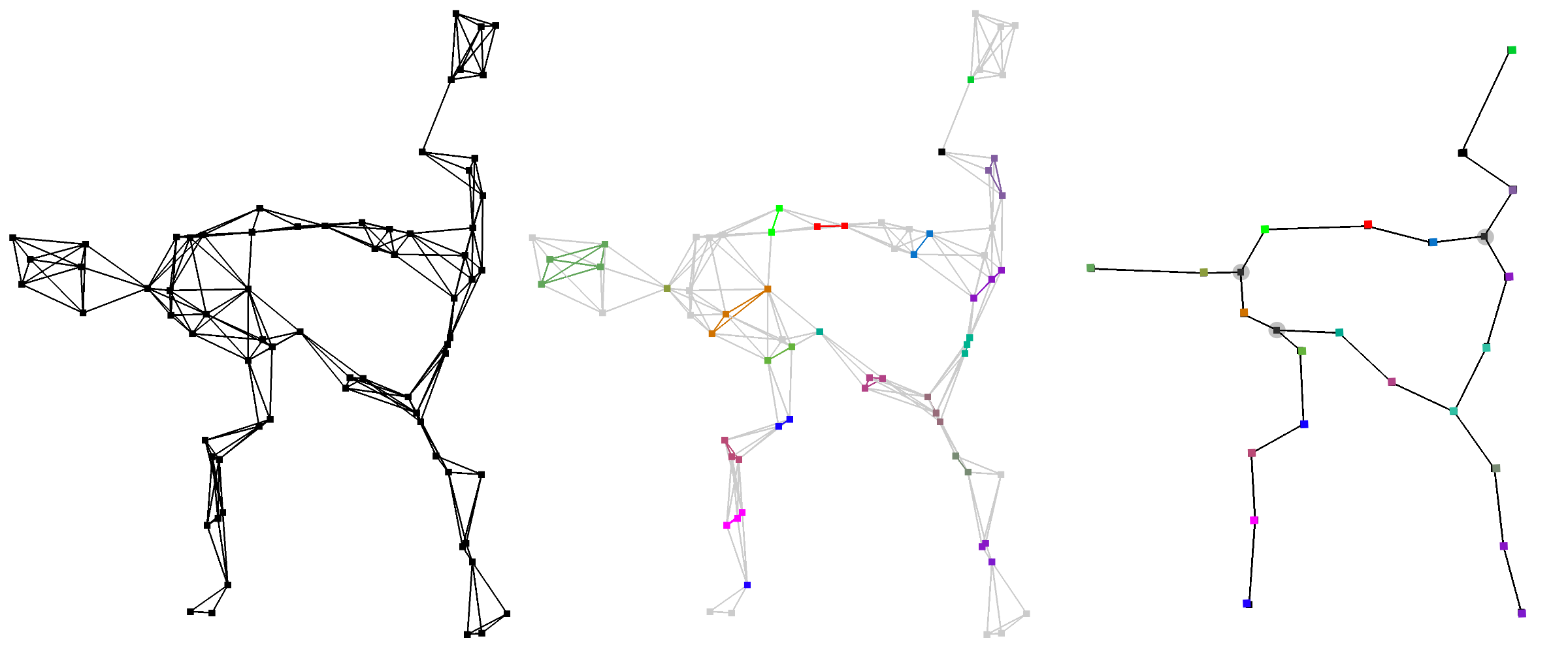}
    \caption{For a simple 2D graph (left), we find the local separators shown as colored vertices and edges (middle). The separators are then replaced with vertices, thereby forming the skeleton (right).}
    \label{fig:cartoon}
\end{figure}

\begin{figure*}[t]
    \centering
    \includegraphics[width=\textwidth]{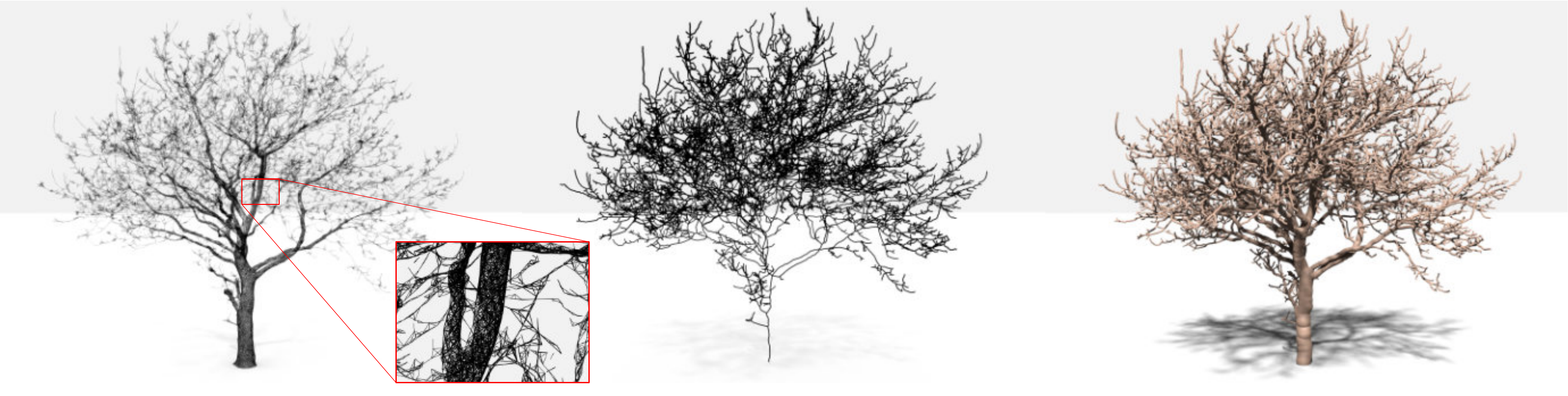}
    \caption{Skeletonization and reconstruction of a botanical tree. The image on the left is the input graph, and the inset shows a close-up where the individual edges are easier to see. The middle image shows the skeleton produced by our algorithm, and the image on the right is a reconstruction of the tree using convolution surfaces.}
    \label{fig:tree-lineup}
\end{figure*}

In this paper, we propose such an algorithm which computes curve skeletons from spatially embedded graphs. This algorithm applies to a range of representations from triangle meshes over voxel grids to graphs constructed from scattered points. Moreover, for a given shape, we may have a different situation in different places: it is quite possible that thin structures in one place are represented by a sequence of connected nodes that already resemble a skeleton, while thicker structures are represented by a different part of the same graph where the points lie in a 2-manifold. A very simple example of our algorithm applied to a 2D toy example is shown in Figure~\ref{fig:cartoon} and Figure~\ref{fig:tree-lineup} shows the result of our algorithm applied to the problem of reconstructing a tree from a point cloud. Initialy, we construct a graph from the point cloud, then we compute a skeleton using the proposed method, and, finally, the skeleton is garbed using convolution surfaces \cite{bloomenthal1991convolution} producing the model shown on the right.
\subsection{Contributions and Overview}
\begin{figure}[h]
    \centering
    \includegraphics[width=0.5\columnwidth]{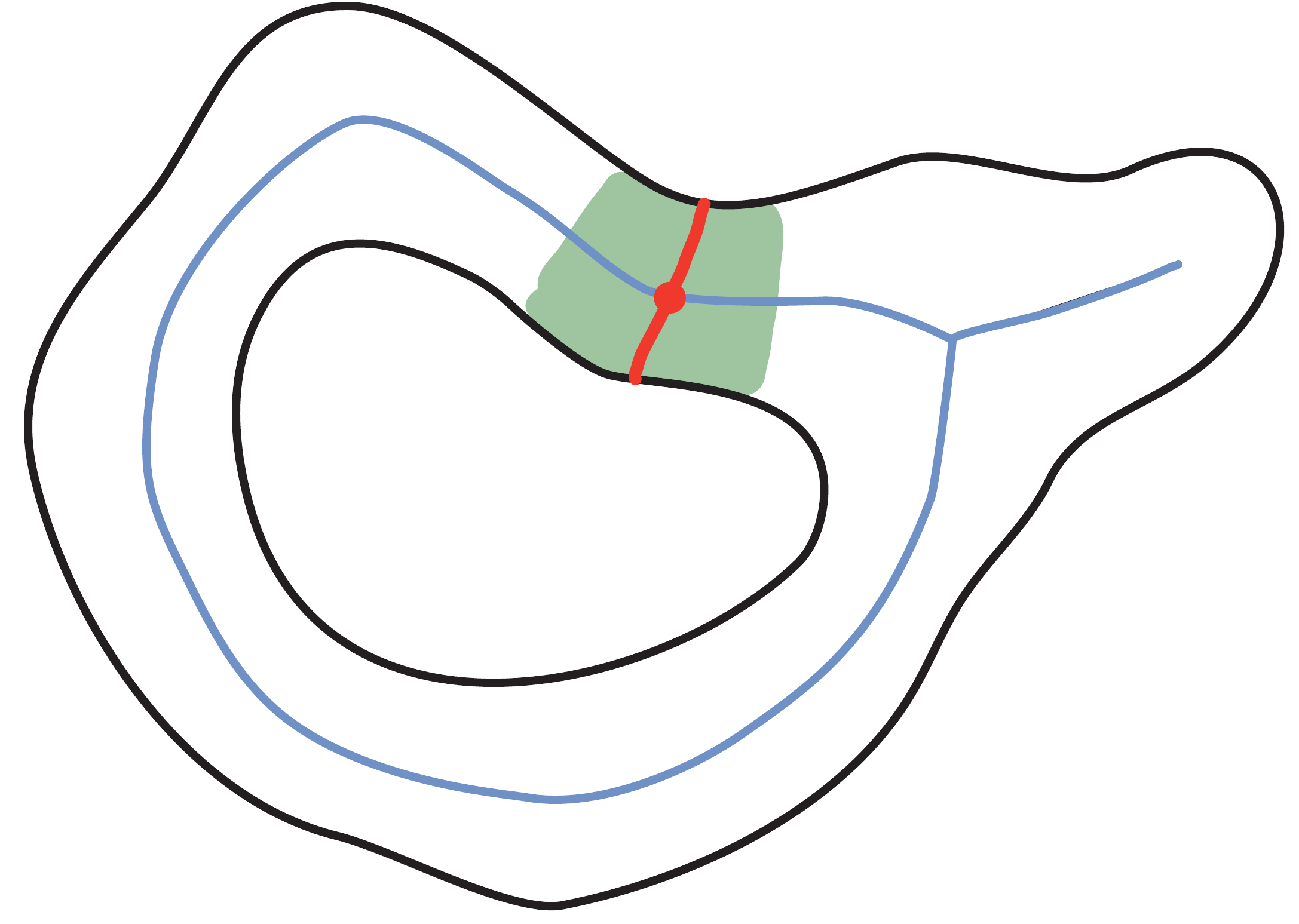}
    \caption{This figure shows a shape (black outline) and its skeleton (blue). The skeletal atom (red curve) contracts to the point on the skeleton shown as a red disk. Note that the skeletal atom is a separator in the part of the shape indicated by the green region.}
    \label{fig:local-separator-continuous}
\end{figure}

Unlike the notion of a \textit{medial surface}, a precise definition of a \textit{curve skeleton} is elusive, but a number of properties are generally agreed upon. In particular, a curve skeleton is understood to be a locally centered shape abstraction such as could be obtained from a given shape by a process of continuous contraction until we arrive at a 1D structure.
Given a point on a curve skeleton, we can identify a set of points on the original shape which contracts to precisely this point on the skeleton. We can think of this set as a \textit{skeletal atom}. Now, if we consider a part of the shape whose corresponding sub-skeleton does not contain any loops, a skeletal atom contained within this part can be construed as a \textit{separator}: its removal will disconnect the part (cf. Figure~\ref{fig:local-separator-continuous}).
We will use the term \textit{local separator} to clarify that a skeletal atom is not necessarily a separator for the entire shape. It should also be emphasized that since our method operates on discrete shapes, we are looking for discrete separators.
Fortunately, the notion of a separator is well known in graph theory, and casting the search for skeletal atoms as a search for vertex separators in a graph is what allows us to generalize the skeletonization process to any object that can be represented as a spatially embedded graph. As mentioned above, this allows us to skeletonize meshes, voxel grids, and graphs constructed by connecting points in space to other points in their vicinity such as the tree shown in Figure~\ref{fig:tree-lineup}. It bears mentioning that this type of data does not define a precise surface, but our method can still compute a skeleton because it is based only on the position and connectivity of the vertices in the input graph.

Normally, the search for skeletal atoms is coupled with the process of finding the skeletal structure itself. For instance, if we find the skeleton by a process of contraction, thinning, or segmentation, the skeletal atoms are generally found on the same cadence as their connectivity. Using the local separator property, we can find skeletal atoms independently of how they are connected. Of course, we also need geometric criteria for whether a separator is a useful skeletal atom, but we can cast a wider net in our search.

Our first contribution is a method which finds local separators in a spatially embedded graph. We propose an algorithm which, starting from a single vertex, grows a connected set of vertices until this set becomes a separator. Subsequently, we shrink the separator again until it becomes a minimal separator. This method is discussed in Section~\ref{sec:local-separator-finding}.

Our second contribution is a procedure for selecting a non-over- lapping subset of the found separators. This is achieved by casting the selection as a weighted set packing problem. The fact that the separators are minimal allows us to pack them far more densely than otherwise, obtaining a skeleton that appears to often resolve comparatively fine details.
The approach is discussed in Section~\ref{sec:local-separator-packing}. 

While our method for finding local separators leads to a good result, it would not be hard to come up with other methods than the one proposed. For instance, methods based on Reeb graphs could be used to generate multiple skeletons which we can see as selections of local separators, and then these skeletons can be blended using the packing method. An example of this is shown in Figure~\ref{fig:reeb-blending}.

From the packed set of separators, we can easily extract a skeleton as discussed in Section~\ref{sec:skeleton-extraction}. The result of local separator finding and packing as well as skeleton extraction is shown in Figure~\ref{fig:cartoon} for a simple 2D input graph.

\subsection{Related Work}
Blum defined the \textit{medial axis} of a given shape as the locus of points where a wave propagating uniformly in all directions from the boundary collides with itself \shortcite{Blum1967}. In 2D this locus is a collection of curves, and in 3D it is a collection of surface components denoted the \textit{medial surface} \cite{siddiqi2008medial}. The medial surface (or axis) can also be defined in a number of other ways, and these several definitions have given rise to a range of methods for computing the medial surfaces of a shape \cite{TagliasacchiSTAR,saha2016survey}. Almost all of the methods are somewhat sensitive to noise, but in recent work Rebain et al. proposed an algorithm that produces approximately inscribed maximal balls from an unorganized surface point cloud \shortcite{rebain2019lsmat}. 

An important property of the medial surface is that it is, in principle, invertible allowing us to reconstruct the shape. However, for many applications, such as creating armatures for animation, reconstruction of botanical objects, or computing shape descriptors, it is of greater utility to obtain a curve skeleton \cite{Cornea07}, and since this is our concern, we will focus on curve skeletons in the following.

There is no universally accepted definition of curve skeletons, but arguing that it should lie in the medial surface, Dey and Sun \shortcite{dey2006defining} proposed an approach where the medial surface is found first, and then the curve skeleton is subsequently found as the set of critical points of the function that measures the distance between the points touched by the maximal ball at each point of the medial surface. This is effective, but requires a well defined surface since the geodesic distance function is used. A faster algorithm which also makes it possible to simplify the medial surface to a curve structure was later proposed by Yan et al. \shortcite{yan16erosion} based on the notion of erosion thickness. While these two works define the skeleton in terms of the medial surface, our method does not rely on the medial surface or require that the input is a manifold surface.

Au et al. \shortcite{au2008skeleton} and later Tagliasacchi et al. \shortcite{Tagliasacchi12Mean} provide algorithms based on mean curvature flow. Tagliasacchi et al. also let the skeleton be attracted by the medial surface. Again, these approaches require a manifold surface mesh. 
A related approach is that of Zhou et al.  \shortcite{zhou2015generalized}, who propose an algorithm for computing the generalized cylinder decomposition of 3D shapes. The cylinder decomposition is highly related to the straight edge skeleton, in the sense that the central axes of cylinders correspond to the edges of skeletons.
Jiang et al. \shortcite{jiang2013curve} follow the strategy of combined clustering (of mesh vertices) and contraction (of skeletal edges) while maintaining a 1-1 relation between clusters and skeletal vertices. Like Jiang et al. we also find a partitioning of the input vertices, but their clustering approach seems to lead to larger clusters and hence a coarser skeleton.
Using machine learning, Xu et al. recently employed volumetric deep learning to infer skeletons for animation from 3D shapes \shortcite{xu2019predicting}.

For volumetric images, there are established methods based on mathematical morphology \cite{serra1983image} for both medial surfaces and skeletal curves \cite{saha2017skeletonization}. An example is the thinning approach by Lee et al. \shortcite{lee1994building} which converts a binary image to a corresponding binary image of the skeleton.

For point clouds, the literature is more sparse. Tagliasacchi et al. find points of approximate rotational symmetry from incomplete point clouds and reconstruct a skeleton from a collection of these \shortcite{Tagliasacchi09}. Huang et al. also compute skeletons from point clouds without connectivity \shortcite{huang2013l1}. Their algorithm initially find medial points which minimize a weighted sum of distances to the input points. A second energy term is used to repel these points in order to avoid clustering. The use of the L1 norm helps reduce the influence of outliers. A notion related to point cloud skeletons is that of an Euclidean Steiner tree~\cite{Jarnik1934,KORTE20011}; i.e. the minimal weight tree spanning the points of the point cloud. Conceivably, a Steiner tree could be pruned to compute the skeleton of a point cloud, but it seems infeasible due to the computational complexity of the approach.

The Reeb graph is an established tool in shape analysis \cite{biasotti2008reeb}. Intuitively, a Reeb graph is constructed by contracting connected components of isocontours of a given height function to a single point. Doing so for the entire shape leads to a structure which is closely related to the notion of a curve skeleton. Tierny et al. find a number of feature points and then use geodesic distance to closest feature point as the height function in a scheme that finds discrete contours based on which they construct a discrete Reeb graph \shortcite{tierny2008enhancing}. These discrete contours could be used as local separators in our scheme, but, unfortunately, Reeb graphs are often suboptimal skeletons, having junctions very close to the surface. In a slightly similar effort, Dey et al. propose a method that finds discrete contours based on Reeb graphs and use it to identify handle and tunnel loops \shortcite{dey13efficient}.

Several authors share our interest in reconstructing botanical trees by first computing a skeleton from images \cite{Quigley18} or point clouds \cite{livny2010automatic,preuksakarn2010reconstructing,gao2019force} and then reconstructing the tree surface from the skeleton.

%


\subsection{Preliminaries}
%
%
%
%
%


%

Since we operate on discrete input, we will assume that our shape has been sampled, producing a set of vertices, $V$, and that a geometric position, $\mathbf p_v$, is associated with each vertex $v \in V$. We will not make any assumptions about whether the vertices are sampled from the surface (i.e. boundary) of the shape or from the interior, but we do require that the vertices are connected by edges, $E$, thus forming a graph $G=<V, E>$. To exemplify, the graph could be an embedded graph as in the case of a manifold triangle mesh, but it could also be a k-nearest neighbor graph defined on a set of scattered points or something else.

%
\subsubsection{Separator and Local Separator}
\label{sec:reconstruction-condition}
Given a graph $G$, a \emph{separator} $\Sigma$ is a subset of vertices with the property that its removal will disconnect the graph into at least two non-empty sets. The size of a separator is simply the number of vertices it contains, and we say that a separator is \textit{minimal} if removing any of its vertices would result in it no longer being a separator.

Say our shape is a cylinder, and let the graph vertices form a regular quad mesh where edges are either parallel to the axis of the cylinder or perpendicular to the axis. Observe that the perpendicular edges form rings around the cylinder, and the vertices connected by these edges separate the vertices on either side of the rings. For a cylinder, it seems natural to define the skeleton by connecting the centers of these rings. Thus, some separators relate to the notion of a skeleton in a useful way, but not all separators are useful. For instance, the neighbors of any vertex is a separator that separates said vertex from the other vertices in the graph, but it is rarely useful. It is also clear that separators do not easily describe skeletons of higher genus surfaces; to overcome this challenge, we turn to local separators.


Given a graph, $G$, an \emph{induced} subgraph $G'$ consists of a subset $V'$ of the vertices $V$ of $G$, and all of the edges in $G$ that connect a pair of vertices in $V'$. Its \emph{boundary} $\partial G'$ consists of those vertices of $G'$ that are linked to the rest of $G$, that is, those that have edges to $V\setminus V'$.
Given a graph, $G$, a \textit{local separator} is a separator, $\Sigma$, of some connected induced subgraph $G' \subseteq G$ with the further property that no boundary vertices belong to the separator, $\partial G' \cap \Sigma = \emptyset$, and that $\partial G'$ is disconnected by the separator. The main advantage of local separators is that they give information about the structure even on higher genus surfaces. As a welcome secondary advantage, they are more computationally efficient to find. As in the case of normal separators, we can very easily find local separators that bear no meaningful relation to the skeleton of a shape. However, given an algorithm that finds appropriate local separators, we can hope to obtain a meaningful skeleton. Such an algorithm will be described in Section~\ref{sec:algorithm}. Note that since we only refer to local separators in the following, the word ``separator'' should be taken to mean ``local separator''.
\subsubsection{Discrete Curve Skeleton}\label{def:skel}
Given a discrete shape represented through a graph, $G$, a \textit{discrete curve skeleton}, $\skel(G)$, is a graph where each vertex either corresponds to a \textit{local separator} of $G$ or is an \textit{auxiliary} branch vertex which joins several vertices. (In a sense, these auxiliary vertices are formed to replace a hyper-edge by a star graph.)

Desirable properties of a skeleton include: (i) It captures the homology of the shape (i.e. it is a deformation retract), (ii) it captures the geometry of the shape, (ii') including all features, while still (ii'') handling noise consistently, and (iii) it is centered in the shape, in particular, (iii') the skeleton bones are contained within the ``meat'' of the shape. We say a skeleton is \emph{good} if it has these properties. If the skeleton successfully lives up to the third criterion, it will be the case that (iii'')
the skeleton is as smooth as the original shape. 

Note that there is a fine balance between being robust against noise, and being able to capture all features.

\subsubsection{Reconstruction Conditions}
Several methods for skeletonization operate by contracting the shape towards the skeleton. This entails the risk that some features may by missed or that noise might be represented as features. It  depends on the speed of contraction, and usually there are operating parameters which the user will have to adjust to stay clear of these issues.
To a large extent, we sidestep the problem by not contracting the shape. Our method operates by finding sets of vertices, i.e. separators, such that any path from the tip to the base of a feature must pass through this cluster. This allows us to facilitate a relatively broad range of inputs as long as they are in the form of spatially embedded graphs. Specifically, 
our algorithm works well both if vertices are sampled only from the surface and if they are also sampled in the interior.

However, 
there are some implicit conditions necessary to 
ensure that we obtain a \emph{good} skeleton
as defined above in Section~\ref{def:skel}.
Note that the number of local separators we can find depends on how connected the graph is. For instance, a complete graph cannot have a separator (local or global) since all vertices are connected. Thus, we need the right amount of edges in order to find a collection of local separators which produces the desired skeleton. 

\begin{figure}[h!]
    \centering
    \includegraphics[width=0.9\columnwidth]{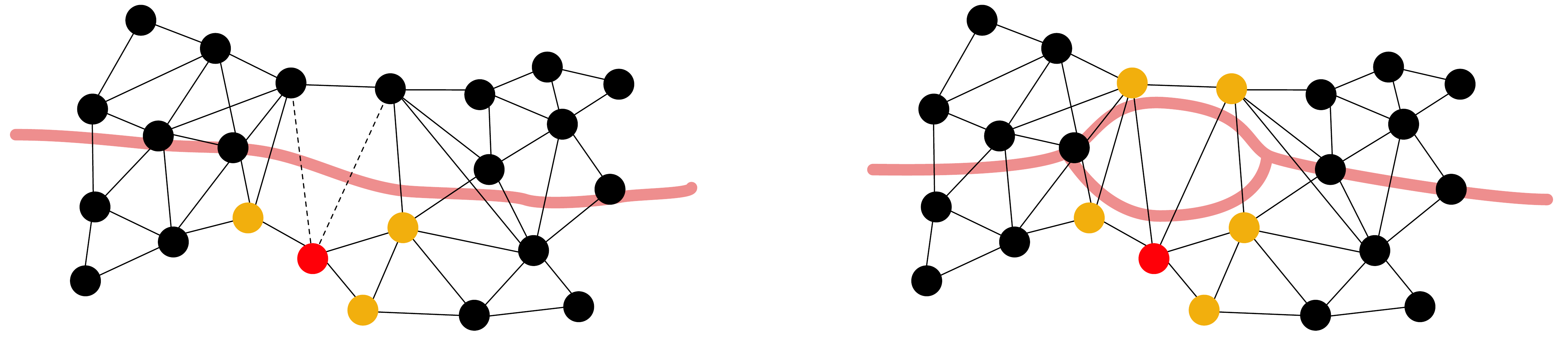}
    \caption{On the left, the skeleton for a subgraph (represented by all the filled circles) is known a priori to contain no cycles (pink curve). Yet, the smaller subgraph (yellow and red vertices) does admit a separator (red) which is not a separator in the containing subgraph. Hence the sufficient connectivity condition is violated unless the edges indicated by dashed lines are included. On the right, the skeleton is known to contain a cycle, but now the red dot doesn't separate any nodes, and the connectivity is excessive.}
    \label{fig:connectivity-criteria}
\end{figure}

This reasoning leads to the following reconstruction condition (illustrated in Figure~\ref{fig:connectivity-criteria}) which assumes that we locally know the (continuous) skeleton and wish to discover whether the connectivity is sufficient to reconstruct (the topology of) this skeleton from the graph.

%
Say we are given a subgraph, $G'$, whose edges and vertices are known to be sampled from a coherent region for which the corresponding part of the (continuous) curve skeleton is known to contain no cycle. If there is a local separator, $\Sigma$, belonging to a smaller subgraph $G'' \subset G'$ then $\Sigma$ must also be a local separator of $G'$. If this is true, \textit{the connectivity is sufficient}, otherwise it is insufficient. 

To understand why this condition must be fulfilled, consider the opposite: if the larger region $G'$ does not have $\Sigma$ as a separator, we can traverse vertices of $G'$ to go from one vertex of $G''$ to another vertex of $G''$ from which the first one would have been separated by $\Sigma$ if the path had been restricted to $G''$, but since the part of the skeleton corresponding to $G'$ is not supposed to contain a cycle, this should not have been possible.

We can use the same condition with different assumptions in order to test for \textit{excessive connectivity}. Assume the part of the shape covered by $G'$ contains precisely one loop, but there is no induced subgraph $G'' \subset G'$ which contains a separator that is not also a separator of $G'$, then the connectivity is excessive because the supposed cycle in the skeleton cannot be recovered by a collection of separators.

%% file: Algorithm.tex
\section{The Local Separator Skeletonization Algorithm}
\label{sec:algorithm}
The  algorithm accepts a graph as input and produces a discrete curve skeleton in the form of another graph as output. Each vertex of the output graph corresponds to a local separator of the input graph. Hence, we can think of the local separators as skeletal atoms, and skeletonization largely becomes the process of searching for a collection of local separators. 

\begin{figure}[h!]
    \centering
    \includegraphics[width=\columnwidth]{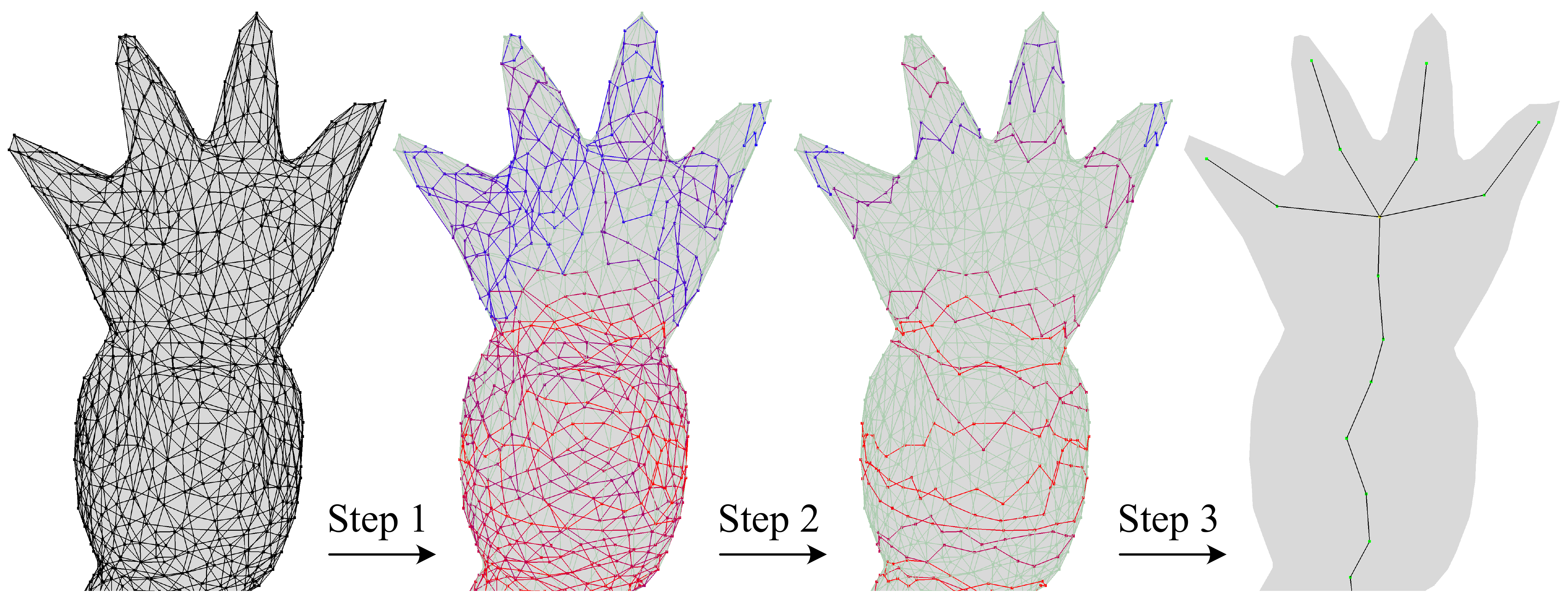}
    \caption{The three steps of the skeletonization process is shown here for a detail (a hand) from the Armadillo model. From left to right the images show the original graph, all the found separators, the packed separators, and the output skeleton. Colors indicate quality (red means high quality, blue means low quality).}
    \label{fig:skeletonization-steps}
\end{figure}

Informally, our method works as follows. Initially, we sample vertices on the input graph and grow a region around each sampled vertex. At the point where the vertices just outside a given region form several disjoint components, that region is a separator which we then shrink back until it becomes minimal. This procedure is described in detail in Section~\ref{sec:local-separator-finding}. Our method for finding separators produces a large set of overlapping separators, and we use set packing to obtain a smaller set of disjoint separators. This is described in Section~\ref{sec:local-separator-packing}. Finally, the skeleton is produced as the quotient graph with respect to a partitioning induced by the separators as we discuss in Section~\ref{sec:skeleton-extraction}. The three steps are also illustrated in Figure~\ref{fig:skeletonization-steps}.
\subsection{Computing a Local Separator}
\label{sec:local-separator-finding}
\begin{figure*}
    \includegraphics[width=\textwidth]{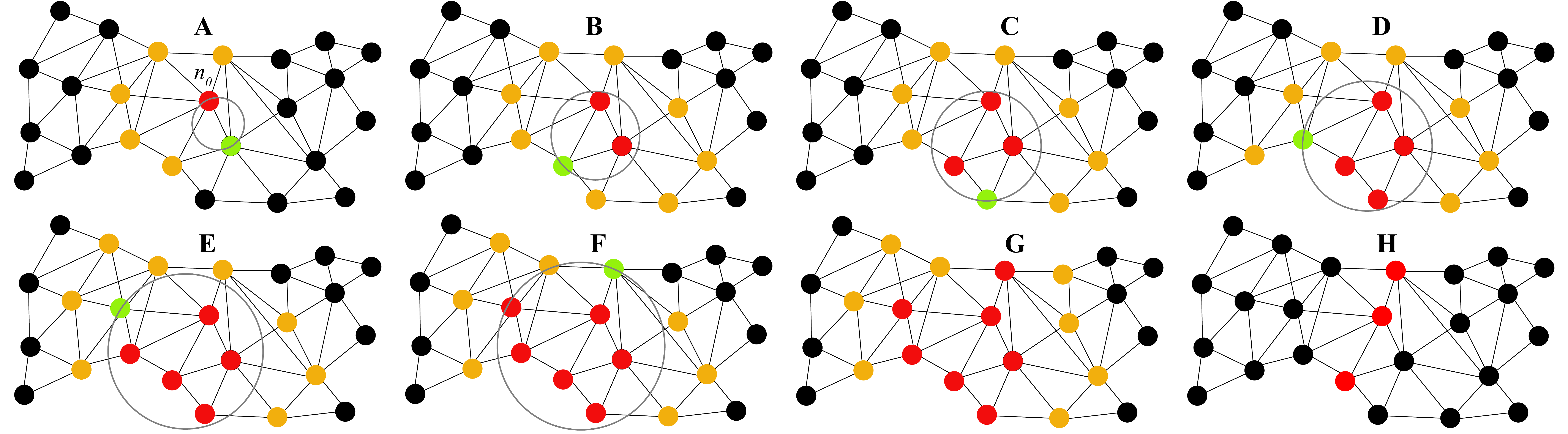}
    \caption{The algorithms for computing a single local separator. A-G illustrate the operation of Algorithm \ref{alg:local-separator}. The color red indicates a separator vertex, yellow a front vertex, and green is used for the front vertex closest to the sphere center Image A shows the situation after we have found the neighbours (yellow and green) of start vertex, $n_0$ (red) and the closest of the neighbors (green). We now use $n_0$ and the closest vertex to redefine the sphere so it precisely contains both vertices. In B we see how all neighbors of the previously green sphere are now added to the front and the sphere redefined again. Images C through G illustrate how more and more vertices are added to both the front and separator until the front is split into two components. At this point the separator is in fact a local separator. Image H shows the separator after it has been thinned using Algorithm \ref{alg:shrinking}.}
    \label{fig:local-separator-computation}
\end{figure*}
The algorithm for computing local separators is based on region growing. We assign an initial vertex to the separator $\Sigma$ and add all its neighbors to the front which we denote $F$. Once a vertex has been added to $\Sigma$ any neighbors of the added vertex which are not in $\Sigma$ are added to $F$ unless they already belong to $F$. Thus, at any time, $F$ is simply the set of vertices not in $\Sigma$ but neighbors to vertices in $\Sigma$.

Instead of growing evenly in all directions, we maintain a bounding sphere around $\Sigma$, and, in each step, we find the vertex in $F$ that is closest to the sphere and add it to $\Sigma$ while expanding the sphere to exactly contain the vertex. It is worth pointing out that the vertex in $F$ closest to the sphere center may be inside the sphere already. In this case the sphere is unchanged. 

The algorithm stops when the front $F$ consists of two (or more) connected components as illustrated in Figure~\ref{fig:local-separator-computation} (G). At this point, $\Sigma$ is a separator of the subgraph $G=\Sigma \bigcup F$, and thus a local separator according to our definition from Section~\ref{sec:reconstruction-condition}, and $\partial G = F$. The benefit of growing the separator using this approach is that it favors going around as opposed to going along features. The center of the expanding sphere that we use to guide the region growing will be used also in the next step when we shrink the separator to a minimal separator, and its radius provides a hint as to local feature size. However, the sphere is not a \textit{minimal} bounding sphere of $\Sigma$ or possessed of other specific properties of which we are aware.

The region growing is illustrated in Figure~\ref{fig:local-separator-computation} (A-G).

\begin{algorithm}
    \begin{codebox}
        \Procname{$\proc{Local-Separator}(G,v_0,\tau)$}
    \zi    $\Sigma \gets \{v_0\}$
    \zi    $F \gets \proc{Neighbors}_G(\Sigma)$ \hspace{2em}\Comment {$F$ is $\partial (G\setminus \Sigma)$}
    \zi    \If $|F| = 0$ \Return $(0,\emptyset)$ 
    \zi    \If $|F| = 1$ \Return $(1,\Sigma)$
    \zi    $C^F \gets  \proc{Connected-Components}_G(F)$
    \zi    $\mathbf c \gets \mathbf p_{v_0}$ \hspace{1em}\Comment{centered in $v_0$}
    \zi    $r \gets 0$ \hspace{1.75em} \Comment{radius $0$}
    \zi    \While $|C^F|==1 \; \vee \; \proc{Front-Size-Ratio}(C^F) < \tau$
    \Do
    \zi         $v \gets \arg\min_{f \in F}\|c-\mathbf p_f\|$ 
    \zi         \If $\|c-\mathbf p_v\|>r$
    \Then
    \zi         $r \gets \frac{1}{2}  (r + \|\mathbf c-\mathbf p_v\|)$ \hspace{1em}
    \zi         $\mathbf c \gets \mathbf p_v + \frac{r}{\epsilon+\|\mathbf c-\mathbf p_v\|} (\mathbf c - \mathbf p_v)$ \hspace{1em}
    \End
    \zi         $\Sigma \gets \Sigma \;\bigcup\; \{ v\}$
    \zi         $F \gets (F \;\bigcup\; \proc{Neighbors}_G(v)) \backslash \Sigma$
    \zi         \If $|F| = 0$ \Return $(0,\emptyset)$
    \zi    $C^F \gets \proc{Connected-Components}_G(F)$
    \End 
    \zi     $q \gets \proc{Front-Size-Ratio}(C^F)$
    \zi $\Sigma, C^F \gets \proc{Shrink-Separator}(G,\Sigma,C^F, \mathbf c)$
    \zi \If $|\proc{Connected-Components}_G(\Sigma)|>1$ \Then \zi \Return $(0,\emptyset)$ \End
    \zi \Return $(q, \Sigma)$
    \end{codebox}
    \caption{Local Separator.\newline
    {\small The arguments are the graph $G$, the initial vertex, $v_0$, and a threshold, $\tau$. $\proc{Front-Size-Ratio}(C^F)$ computes the ratio of the cardinalities of the smallest set to the largest set in a collection of sets passed to the function. $\proc{Connected-Components(G,F)}$ computes the connected components of the vertex set $F$ given the connectivity of the graph $G$. $\epsilon$ is a small constant preventing division by zero.}}
    \label{alg:local-separator}
\end{algorithm}
If the algorithm is invoked for a valence 1 vertex (i.e. leaf), it immediately returns a separator consisting only of this vertex. Thus, leaf vertices are defined to be local separators. This is because the skeletonization is designed to be \textit{idempotent}: applying skeletonization to a skeleton should not change it. Since the interior vertices of a skeleton are separators unless the skeleton contains cliques, this property can be attained simply by defining leaves to be separators. The algorithm also stops and returns an empty separator if the front is empty at any point since, in this case, we have flooded an entire connected component. 

If neither condition is met, the algorithm continues while $F$ consists of a single significant connected component. In this context, ``significant'' means that the we also continue if the size of the smallest component is very small compared to the largest. Having found two significant components, the algorithm stops, and $\Sigma$ is first shrunk to a minimal separator (using an approach described below) and then returned, unless the shrunk $\Sigma$ consists of more than a single connected component. If this happens, we return an empty separator. Before returning a non-empty separator, we compute the quality which is simply the ratio of the smallest front component of $F$ to the largest. Pseudo-code for $\proc{Local-Separator}$ is shown in Algorithm~\ref{alg:local-separator}.


\subsubsection{Shrinking a Separator}
\begin{algorithm}
    \begin{codebox}
    \Procname{$\proc{Shrink-Separator}(G,\Sigma, C^F, \mathbf c)$}
    \zi $\mathbf P \gets \proc{Laplacian-Smooth}(G,\Sigma)$
    \zi \For $v \in \Sigma$ \Do
    \zi     $d_v \gets \|\mathbf P_v - \mathbf c\|$
        \End
    \zi \Repeat
    \zi     $v \gets \argmax_{\nu \in \Sigma} d_{\nu}$ 
    \zi 	$\gamma = \emptyset$
    \zi     \For $w \in \proc{Neighbors}_G(v)$ \Do
    \zi         $\gamma \gets \gamma \cup\{ i \;|\; w \in C^F_i \}$ \End
    \zi         \If $|\gamma| = 1$ \Then
    \zi             $\Sigma \gets \Sigma \backslash v$
    \zi             $C^F_i \gets C^F_i \cup \{v\}$ where $\{i\}=\gamma$
               
            \End
        \End
    \zi \Until $\proc{Is-Minimal-Separator}(G, \Sigma)$
    \End
    \zi \Return $\Sigma, C^F$
    \end{codebox}
    \caption{Shrink Separator.\newline
    {\small The arguments are the graph $G$, the separator, $\Sigma$, and the front components $C^F = \{C^F_1, C^F_2, \ldots\}$. $\proc{Laplacian-Smooth}(G,\Sigma)$ computes the Laplacian smoothing of the vertices that belong to $\Sigma$ keeping the vertices in the rest of $G$ fixed and returns the smoothed positions.}}
    \label{alg:shrinking}
\end{algorithm}
We want to find a separator that is minimal, but also smooth and balanced. The separator minimization is performed by iteratively removing vertices from the separator until no vertex can be removed. The result clearly depends on the order in which we remove vertices from the separator.

A simple heuristic would be to remove vertices in order of decreasing distance to the center of the sphere used to find the separator in the first place (as discussed above). While this is effective, it has the deficiency that for irregular structures, this can lead to somewhat meandering separators. We initially perform Laplacian smoothing of the initial separator, $\Sigma$, using inverse edge length as the weighting scheme and with fixed positions for the vertices of $F$. Having smoothed the separator, we remove vertices in order of decreasing distance to the sphere center until the separator is minimal. There are a few cases where the shrunk separator breaks into disjoint components. We test for and discard these results.
Pseudo-code for the function, $\proc{Shrink-Separator}$ is shown in Algorithm~\ref{alg:shrinking}.
\subsubsection{Optimizing Minimal Separators}
\label{sec:optimizing-minimal-separators}
A minimal separator is optimal in the sense that no vertex can be removed from it without it ceasing to be a separator. However, we can also measure other qualities in a separator such as the sum of the length of graph edges connecting two vertices that both belong to the separator. Below, we describe an optional step that can be used to optimize minimal separators in this sense.

It is possible to create a variation of a minimal separator, $\Sigma$, by exchanging a vertex, $v \in \Sigma$, with a neighbor vertex, $u \notin \Sigma$, as long as
\begin{itemize}
    \item $v$ does not have other neighbors that belong to the same front component as $u$, and
    \item $\mathrm N(u) \bigcap \Sigma = \mathrm N(v) \bigcap \Sigma$.
    \item $u$ belongs to the original (unminimized separator)
\end{itemize}
The first of these three conditions ensures that we do not introduce tunnels through the separator. The second condition ensures that $u$ has the same neighbors in $\Sigma$ as $v$. This translates into a more gradual change to the separator which we have observed leads to better results. The final condition ensures that we do not include vertices from outside the subgraph created by the search in Algorithm~\ref{alg:local-separator} (before Algorithm~\ref{alg:shrinking}) as this could invalidate the separator.

Abiding by the rules above, we can now perform substitutions in order to minimize some particular measure such as the summed length of edges belonging to the separator, i.e 
\begin{equation}
E(\Sigma) = 0.5 \sum_{u\in\Sigma} \sum_{v \in \mathrm N(u)} \| v - u\| \cdot \left\{ 
\begin{array}{rcl}
    1 & \text{if} &v \in \Sigma\\
    0 & \multicolumn{2}{l}{\text{otherwise}}
\end{array}    
\right.
\label{eq:separator-opt}
\end{equation}
We implemented an algorithm that optimizes a separator by performing any substitution that reduces \eqref{eq:separator-opt} until no further substitutions will reduce the energy. 

This algorithm is applied once to all minimal separators. If it does improve the separator, we perturb it slightly by making random substitutions and then run the optimization again. This procedure is iterated a fixed, small number of times. Each time, we backtrack if the procedure does not improve the separator. 

\subsubsection{Run-Time Analysis}
The time consumption for finding a separator can be analyzed in two terms. Let $s$ be the total number of vertices in $\Sigma$ returned by Algorithm~\ref{alg:local-separator}. First, we spend $s$ steps adding one vertex at a time to the separator. 
In each step, we spend time proportional to the current size of the front $F$ performing a linear scan for the closest vertex to the current center, and furthermore, time proportional to the number of edges between vertices of the front finding the connected components. 
Then, we spend $O(s)$ time per round of diffusion, for a total of roughly $O(\sqrt{s})$ rounds. Thus, the time consumption becomes $O(s\cdot(V\left[F_{\max} \right] + E\left[F_{\max} \right]) + s\sqrt s)$, where $V[G']$ and $E[G']$ are the number of vertices and, respectively, edges of the induced subgraph $G'$, and $F_{\max}$ is the largest front through the course of the algorithm. Thus, in general and in worst-case, we can make no better analysis than $O(s^3)$, since the front may be proportional to the separator, and may have many edges. However, we do assume that the points are sampled somewhat evenly from a surface or a shape in 3D space. If the points are sampled somewhat evenly from a surface, the front size is expected to be $\sqrt{s}$, while if points stem from a shape, the front size is expected to be $s^{2/3}$. Furthermore, if points stem from a mesh or a voxel grid, the number of edges in the subgraph induced by the front is indeed proportional to the number of vertices; indeed, not only will the subgraph be planar, but voxel grids are constant degree by construction. Thus, for meshes, we expect a running time of $s\sqrt{s}$ to find one separator, and 
for volumetric graphs, we expect a running time of $O(s^{5/3})$. For point clouds where vertices have bounded degree, the running time will correspond to that of meshes or voxels, depending on whether the points are sampled from the surface or the iterior of the shape. 

%
%
\subsubsection{Sampling}
\label{sec:sampling}
Since the algorithm for finding a local separator starts from a single vertex of $G$, we need to first choose a set of initial vertices. Clearly, one option is to start the search from \emph{all} vertices of $G$, but we can improve performance dramatically by sampling. However, simply choosing a random subset of the vertices would lead to small features being missed. Broadly speaking, we want many separators to choose from in order to find a dense packing later.

In our experience, a good compromise is to visit all vertices in random order and the use vertex $v$ as starting point for Algorithm~\ref{alg:local-separator} with probability $p(v) = 2^{-\val{v}}$ where $\val{v}$ is the number of times $v$ has been included in a separator. This heuristic prioritizes the more sparsely covered regions, but, of course, it does not prevent that some vertices could be sampled repeatedly: that the separator found starting from $x$ contains $y$ does not imply the converse; for some inputs there may be ``popular'' vertices that belong to many separators even if they are not used as starting point many times.

The total running time of the separator finding step thus becomes the product of the time to find a separator with the number of vertices sampled: $\sum_{v} p(v) \cdot f(s_v)$, where $p(v)$ is the sampling probability for $v$, $s_v$ is the separator ball found from $v$, and $f(s)$ is the time consumption to find the separator which is a function of the data set type (be it voxel, mesh, or point cloud graph). Intuitively, especially for locally cylindrical shapes, the probability of sampling $n$ should decrease as the size of the local separator through $n$ increases, leading to a better total running time.
%
%
\subsection{Packing Local Separators}
\label{sec:local-separator-packing}
The local separators found using the above method correspond, roughly, to cross sections of the shape. It is obvious to consider each separator to be a skeletal atom and map it to a vertex of a discrete curve skeleton. However, the separators overlap which makes it difficult to infer a reasonable connectivity for such a skeleton. 
Our solution is to pack the separators on the graph. In a packing, each vertex can be covered by only a single separator. Thus, we obtain a partitioning of the vertices in the graph into non-overlapping sets - plus an additional set of unassigned vertices, but these can be trivially assigned to the closest separator. From this partitioning, we can easily extract a skeleton using a method described in Section~\ref{sec:skeleton-extraction}.

Set packing is itself an NP-hard problem: Deciding whether some given number of sets from a family exist that cover the entire universe, or optimizing for how few are needed.
Even for subsets of size $k$, no better guarantee than a $k/2$-approximation is known~\cite{Hurkens1989}, and this is tight up to $\log k$-factors~\cite{Hazan2006}. However, for our purposes, a greedy heuristic often yields good results in practice.

We use a greedy, weighted packing scheme \cite{kordalewski2013new} since this allows us to prioritize the best separators. While several weighting schemes were considered~\cite{chvatal1979greedy,kordalewski2013new}, we have chosen a weight based on how balanced the local separator is. Specifically, once the separators have been minimized using Algorithm~\ref{alg:shrinking}, we compute the ratio of the smallest to the largest front component meaning that balanced separators are prioritized. The weight is returned from Algorithm~\ref{alg:local-separator} together with the separator itself.

The heuristic takes $l$ subsets (in our case, separators) of size $S$ and does the following three steps: First, it computes a weighted redundancy count for each separator: how many other separator does it intersect and by how much. Then, it sorts them according to redundancy. Finally, it greedily adds sets of increasing redundancy. The running time is thus $O(l^2 S + l \log l + l^2) = O(l^2 S)$, that is, dominated by the first term corresponding to the first step. 
As future work, we could parallelize the computation behind the first term, leading to an $O(l^2 S/p)$ time algorithm assuming $p$ processors. Also, one could use a MinHash algorithm to estimate set resemblance~\cite{Broder1997resemblance}, in order to speed up computation time for the $\texttt{set}\_\texttt{intersection}$ subroutine.

In our case, we will have at most $n$ different separators, one for each vertex of the graph, of sizes that are worst-case $n$, leading to an $O(n^3)$ running time. However, in practice, for evenly spaced sample points along a surface or within a shape, the smallest separators through a point would be $O(\sqrt{n})$ (assuming bounded genus), and even less for lanky shapes such as a botanical tree (Figure~\ref{fig:tree}) or foam (Figure~\ref{fig:tomofoam}).

\subsection{Skeleton Extraction}
\label{sec:skeleton-extraction}
\begin{figure}[h!]
    \centering
    \includegraphics[width=\columnwidth]{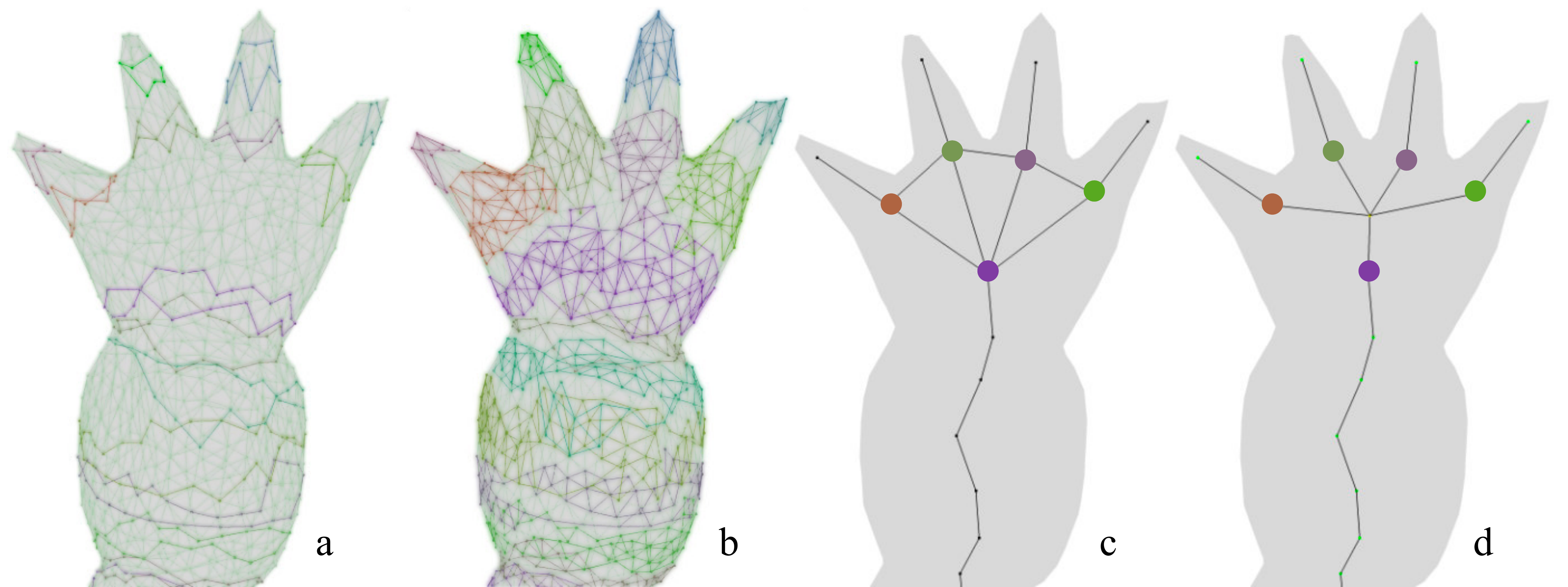}
    \caption{Steps of the skeleton extraction. The original (minimal) local separators (a), the maximized separators (b), the quotient graph (c), and the output (d) where each clique complex (bold colored dots) is replaced by a star graph with the cluster vertices as leaves, and a new vertex as center.}
    \label{fig:skeleton-from-separators}
\end{figure}
From a packing of local separators it is now straight forward to compute the actual skeleton. 
Initially, we \textit{maximize} the separators by assigning all vertices that do not already belong to a separator to their closest separator in terms of distance along the graph edges. Since all vertices are now assigned to a separator, the separators form a partitioning of the graph, and the skeleton is computed as the \textit{quotient graph} by the equivalence relation that two vertices belong to the same maximized separator \cite{sanders2012high}. The position of each vertex of the quotient graph is simply the average of the positions of all of the vertices in the associated separator (i.e. equivalence class). 

The quotient graph is close to the final skeleton, but it contains cliques of size $\ge 3$ and complexes of such cliques. The cliques arise at junctions where several branches meet, leading to several vertices in the quotient graph that are all connected as exemplified in Figure~\ref{fig:skeletonization-steps} (b,c). Often these cliques are of size three meaning that the separators on three branches are all connected to each other. Moreover, in a complex junction where more than three branches meet, the cliques will share sub-cliques, thus forming complexes. For instance, in Figure~\ref{fig:skeletonization-steps} (b) we observe that there are separators on each finger and on the hand. In the quotient graph (c), the separators are now vertices, and the connectivity of the vertices reflects the connectivity of the separators, forming a complex of three cliques.

The final step of skeletonization consists of removing clique complexes from the quotient graph. To do so, we find all cliques of size three. Next, we compute the pairwise intersections between all cliques, merging those which share a clique of size two. All edges in the resulting complexes are removed, and the complex is replaced with a single vertex which is then connected to all vertices of the complex. An example is shown in Figure~\ref{fig:skeleton-from-separators} (d). 

Clearly, the search for $3$-cliques dominates the running time for skeleton extraction, and thus, it is quadratic in the number of packed separators, or, equivalently, quadratic in the number of vertices in the final skeleton.
\subsection{Skeleton Smoothing}
\label{sec:skeleton-smoothing}
In some cases, it is important to get a smooth skeleton, and our scheme has no inherent smoothness since the position of each skeletal vertex is computed independently of other skeletal vertices as the average position of the associated separator vertices. Figure~\ref{fig:smoothness} (a \& b) show examples where a very smooth skeleton results, but the leftmost images of Figure~\ref{fig:smoothness} (c \& d) show two cases where the resulting skeleton is less smooth.

To (optionally) smooth the output skeleton, we use a simple weighted Laplacian smoothing scheme
\begin{equation}
    \mathbf p^{\text{new}}_v = \frac{\mathbf p_v}{|\mathrm N(v)|}  + \left(1-\frac{1}{|\mathrm N(v)|}\right)
    \frac{\sum_{u\in \mathrm N(v)} \frac{\sqrt{W_u}}{\mathrm N(u)} \mathbf p_u}{\sum_{u\in \mathrm N(v)} \frac{\sqrt{W_u}}{\mathrm N(u)}} \enspace ,
\end{equation} 
where $\mathbf p_v$ is the position of vertex $v$, $\mathrm N(v)$ is the set of neighbors, and $W_v$ is the weight. The weight is given by the number of vertices in the separator associated with $v$ in the input graph. If $v$ is a branch vertex that replaces a clique, $W_v$ is simply the average of the weights of the vertices in the clique.

The logic behind this scheme is that we want high weight vertices to attract more but also lower valency vertices since leaf vertices in particular should not be smoothed too much as that could lead to a loss of fine detail. Clearly, this smoothing scheme can be applied iteratively, depending on the desired level of smoothness as illustrated in Figure~\ref{fig:smoothness} (c \& d).

%% file: Testing.tex
\section{Results and Discussion}
We tested our approach on three different types of input: 2-manifold polygonal meshes, volumetric grids and point data. While the algorithm runs unchanged on all three, some preprocessing is required for other inputs than triangle meshes. In the following, we first describe some implementation details and then relate the results of our experiments.

\subsection{Implementation}
The graph skeletonization algorithm is implemented in C++. Graph and mesh data structures are based on our open source XX library (omitted for anonymity). The graph data structure is implemented using adjacency lists.

To speed up computations, we parallelize the sampling process using the threading facilities of C++. Since the decision of whether to compute a separator from a given vertex, $v$, depends on how many separators already include $v$ that decision consequently depends on thread scheduling. This entails that the program is not fully deterministic; there is a slight variation between runs. However, in all cases, the results shown are those produced by a single run of the algorithm - not the best result from several runs.

Our method relies only on one parameter: the quality threshold, $\tau$, in Algorithm~\ref{alg:local-separator} which was set to the empirically chosen 0.0875 in all experiments. Sampling was used in separator finding in  all cases except Figure~\ref{fig:cartoon}.

In our comparisons with other methods, we used the CGAL implementation of Mean Curvature Skeletons \cite{Tagliasacchi12Mean} and the binary kindly provided by the authors to compare with L-1 Medial Skeletons \cite{huang2013l1}.

Except where noted, tests were run on a late 2018 Mac Mini (3.2 GHz Intel Core i7-8700B) running MacOS Catalina.
\subsection{Surface Meshes}
\label{sec:results-surface-meshes}
\begin{table*}
    \small
    \centering
    \begin{tabular}{r|r|rrr|rrr|rrr|rrr}
        & \multicolumn{1}{c}{Input}  & \multicolumn{3}{c}{Skeleton} &  \multicolumn{3}{c}{Leaf}  & \multicolumn{3}{c}{Branch} &\multicolumn{3}{c}{Run Time}\\
        &  \multicolumn{1}{c}{Vertices} & \multicolumn{3}{c}{Vertices} &  \multicolumn{3}{c}{Vertices}  & \multicolumn{3}{c}{Vertices} & \multicolumn{3}{c}{(secs)}\\
\hline

        && LS & \multicolumn{2}{c|}{MCF} & LS & \multicolumn{2}{c|}{MCF} & LS & \multicolumn{2}{c|}{MCF} & LS & \multicolumn{2}{c}{MCF} \\
        &&  & 0.1 & 1.0 &  & 0.1 & 1.0 &  & 0.1 & 1.0 &  & 0.1 & 1.0 \\
\hline

        Human hand & 12438 & 120 & 349 & 271 & 6 & 6 & 6 & 3 & 4 & 4 & 87.82 & 14.97 & 84.00 \\
        M18020 & 75120 & 2719 & 2973 & 3510 & 0 & 2 & 2 & 40 & 80 & 67 & 24.20 & 51.49 & 60.43 \\
        Fertility & 24994 & 181 & 523 & 483 & 3 & 0 & 0 & 7 & 6 & 5 & 534.18 & 38.26 & 229.56 \\
        Wood Sculpture & 19803 & 241 & 501 & 426 & 7 & 5 & 3 & 11 & 10 & 9 & 141.15 & 22.90 & 140.86 \\
        Asian Dragon & 47701 & 1276 & 838 & 720 & 83 & 35 & 29 & 29 & 33 & 27 & 512.39 & 44.04 & 274.99 \\
        Neptune & 28052 & 491 & 606 & 569 & 27 & 9 & 12 & 22 & 11 & 16 & 247.42 & 27.82 & 141.26 \\
        Rocker Arm & 10803 & 67 & 230 & 183 & 4 & 2 & 2 & 4 & 2 & 2 & 107.33 & 19.01 & 86.76 \\
        M22081 & 3970 & 130 & 301 & 290 & 4 & 4 & 3 & 10 & 11 & 11 & 5.92 & 4.18 & 4.50 \\
        Frog & 37225 & 483 & 820 & 712 & 23 & 23 & 22 & 14 & 20 & 18 & 1394.70 & 57.64 & 388.14 \\
        Spider & 4675 & 212 &  &  & 29 &  &  & 1 &  &  & 1.71 &  &  \\
        Noisy Dinosaur & 23302 & 240 & 487 &  & 13 & 34 &  & 7 & 4 &  & 299.63 & 26.78 &  \\
        M21362 & 1828 & 342 & 4 & 36 & 23 & 1 & 8 & 1 & 1 & 3 & 8.76 & 2.65 & 29.77 \\
        Armadillo & 6488 & 270 & 402 & 372 & 20 & 10 & 14 & 9 & 7 & 12 & 11.45 & 9.28 & 67.07 \\
        idem (noisy) & 6488 & 315 & 347 & 346 & 21 & 8 & 13 & 8 & 5 & 8 & 13.61 & 9.25 & 74.22 \\
        idem (noisier) & 6488 & 494 & 348 & 279 & 36 & 8 & 10 & 11 & 6 & 6 & 17.61 & 9.04 & 73.02 \\
    \end{tabular}
    \caption{Summary of the results of triangle mesh skeletonization using our method and the mean curvature \cite{Tagliasacchi12Mean} method (MCF) as implemented in CGAL. We compared MCF using $w_H=0.1$ (default) and $w_H=1$.}
    \label{tab:mesh-comparison}
\end{table*}
\begin{figure*}[p]
    \centering
    \includegraphics[width=0.85\textwidth]{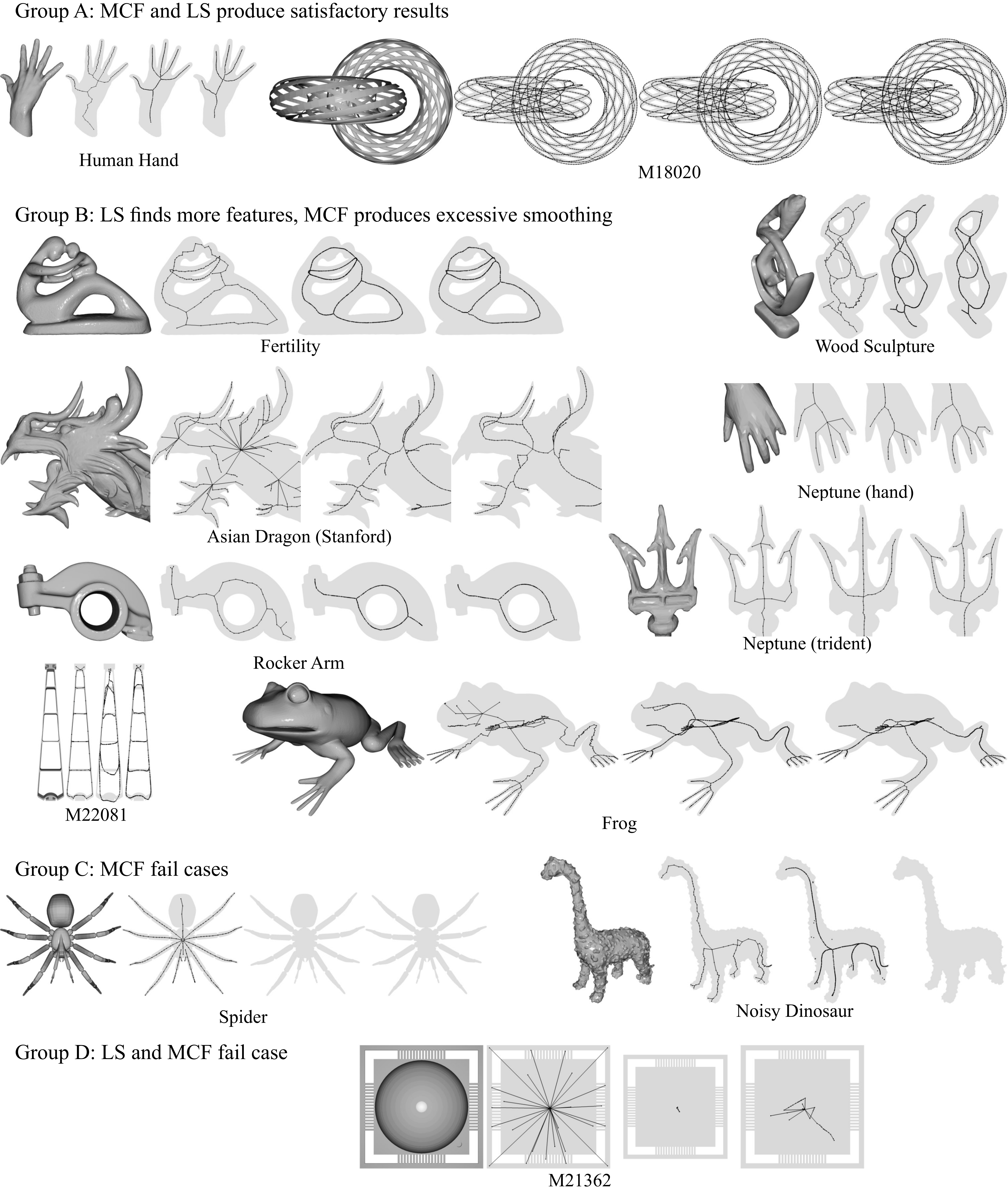}
    \caption{Comparison of triangle mesh skeletonization using our method and the mean curvature  \cite{Tagliasacchi12Mean} method (MCF) as implemented in CGAL. Each mesh is shown shaded on the left, skeletonized with our method (center left) and using MCF on the right (with $w_H=0.1$ center right and $w_H=1$ far right). The models are divided into Groups A through D depending on the outcome for MCF and LS.}
    \label{fig:mesh-comparison}
\end{figure*}
We compared our method (LS) to mean curvature (MCF) skeletonization \cite{Tagliasacchi12Mean} using the CGAL implementation. The method exposes the $w_H$ parameter which ``controls the velocity of movement and approximation quality'' according to the CGAL documentation. We ran our experiments with $w_H=0.1$ (default) and with $w_H=1$ which is slower, but, in some cases, improves quality. The comparison was performed on a selection of  meshes (many commonly used) of varying complexity, genus, and level of geometric detail.  The results are summarized in Table~\ref{tab:mesh-comparison}. 

\begin{figure*}[h!]
    \centering
    \includegraphics[width=\textwidth]{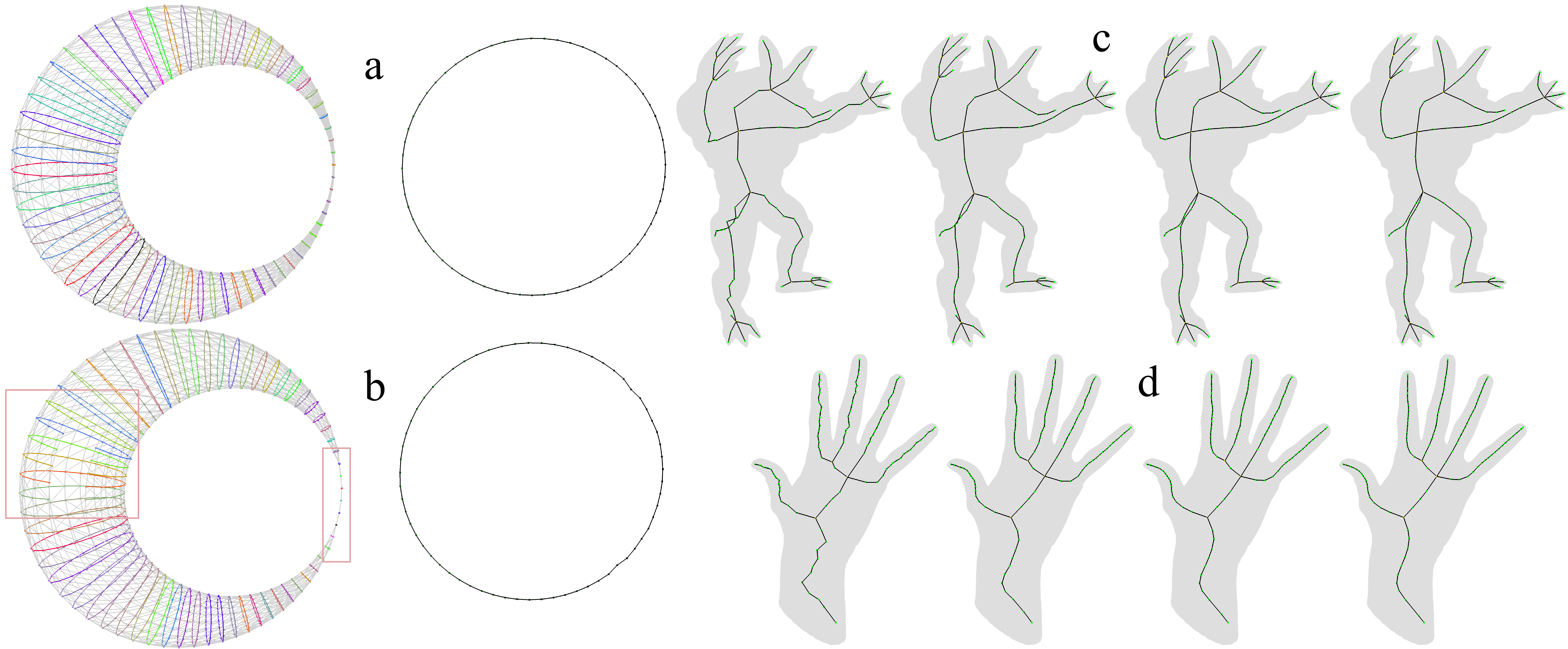}
    \caption{a and b show a Dupin cyclide skeletonized using the LS method. In b, a segment of the surface has been removed introducing a boundary (indicated by a red box) and edge contraction has resulted in the model being collapsed to a curve (indicated by the other red box). In both a and b, the graph of the cyclide is shown left with separators color coded, and the resulting skeleton is shown on the right. c and d show the result of a skeletonization of the armadillo (c) and the human hand (d) with 0, 1, 2, and 3 steps of smoothing going fron left to right.}
    \label{fig:smoothness}
\end{figure*}

\paragraph{General Comparison}
A number of the output skeletons are shown in Figure~\ref{fig:mesh-comparison}, and we have divided the results into four groups which are discussed below. We have assessed them against the criteria sketched in Section~\ref{def:skel}, and against the additional criterion that some skeleton should be computed.
\begin{description}
    \item[Group A: both win] contains the human hand and a synthetic mesh. For these two meshes the LS and MCF results are comparable. It may be observed that the MCF result is smoother in the case of the human hand. However, while smoothing is an intrinsic (and not always desired) part of the MCF method, we can easily obtain a smoother output from the LS method using the approach described in Section~\ref{sec:skeleton-smoothing}, and the result of smoothing the human hand is shown in Figure~\ref{fig:smoothness}(d). 
    \item[Group B: excessive smoothing] contains several cases where the LS method yields a skeleton that captures more details (criterion ii) and/or with higher fidelity than MCF (criterion iii). In the case of Fertility, Rocker Arm, and Wood Sculpture, the difference is mainly that the LS approach captures small protrusions not captured by the MCF skeleton. 
    In the case of the Asian Dragon (only head shown), significant head appendages are missing from the MCF skeleton. When it comes to the frog, the LS skeleton captures eyes and nostrils but both are missing from the MCF skeleton for $w_H=1$ and the eyes are missing for $w_H=0.1$.
    
    The results for M22081 show clearly that too much smoothing can lead to loss of precision. The highly regular structure of M22081 is captured quite well by the LS method whereas the horizontal parts are deformed by the MCF method: excessively so for $w_H=0.1$ and somewhat less for $w_H=1$.  In a similar vein, the overall shape of the trident from the Neptune model is captured better by the LS Method, and MCF misses one or more barbs. In the case of the hand of the Neptune statue, we also observe that the skeleton seems to be over-smoothed.
    \item[Group C: MCF fails] contains two models where MCF fails to produce an output. The spider model consists of multiple intersecting parts and it contains non-manifold edges which is likely the reason for this failure. The noisy Dinosaur contains a few tiny connected components which may both be the cause of the strange isolated components in the skeleton for $w_H=0.1$ and also the reason for the failure in the case of $w_H=1$. We note that the LS method produces valid and reasonable output in both cases.
    \item[Group D: both fail] consists of a single model, M21362, which consists of large, planar regions defined by very few vertices and exceedingly ill-shaped triangles. Both LS and MCF produce a valid but not a reasonable output according to our criterion (i) that the skeleton should capture the homology of the shape, and probably models like this one can only be handled by resampling the mesh.
\end{description}
\paragraph{Comparison of Surface Reconstructions}
\begin{figure*}[h!]
    \centering
    \includegraphics[width=\textwidth]{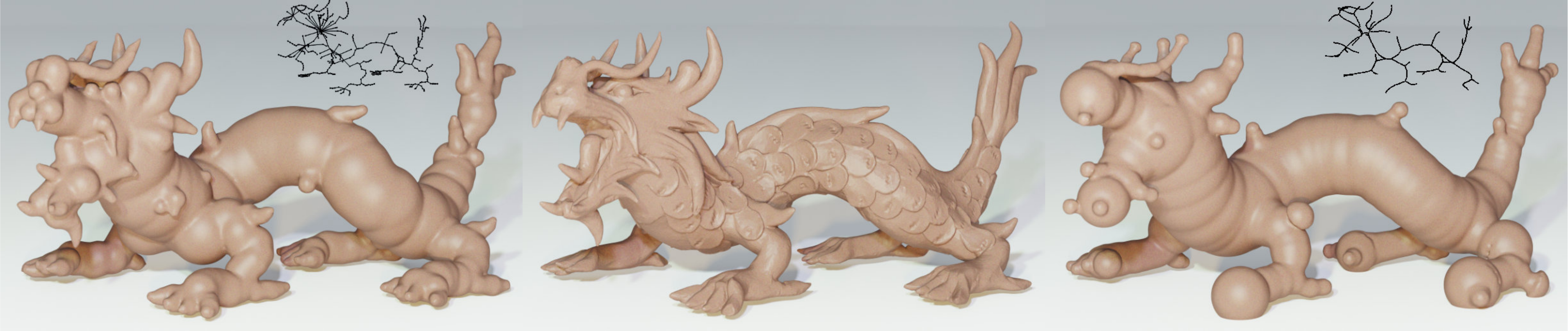}
    \caption{The original triangle mesh of the Asian Dragon is shown in the center. On the left is a surface reconstructed from our LS skeleton using convolution surfaces and on the right a similar reconstruction from the MCF skeleton. In both cases the skeleton used for reconstruction is shown above the rendered model.}
    \label{fig:dragon_comparison}
\end{figure*}
In order to investigate the difference between LS (our method) and MCF, we applied convolution surface based reconstruction \cite{bloomenthal1991convolution} to the output from both methods. Since curve skeletonization is not invertible, this will not result in an accurate reconstruction, but it does facilitate a visual comparison of how well each method qualitatively captures the input surface. We chose to do this on the Asian Dragon since this model has numerous fangs, digits and assorted appendages which provide an interesting challenge. To reconstruct a shape from the skeleton, we associate a radius with each node of the skeleton. This radius is computed as the average of the distances from the skeletal node to the associated vertices. In the case of LS, the associated vertices are of course the separator, and MCF reports the vertices that are contracted to each node. The result is shown in Figure~\ref{fig:dragon_comparison}. It is very clear from the images, that LS captures far more features than MCF: in fact, all digits are captured by the LS skeleton and just a few by the MCF skeleton, and a number of other appendages are also missing from the MCF skeleton. This observation is supported by the data in Table~\ref{tab:mesh-comparison} where we observe that the LS skeleton contains between two and three times more leaf vertices than the MCF skeleton. Because the leaf nodes of the skeleton represent a larger part of the input mesh in the case of the MCF reconstruction, the tips of features appear comparatively bulbous in the reconstruction as seen in Figure~\ref{fig:dragon_comparison}. 
\paragraph{Smoothness of Skeletons}
As previously noted, the result of the LS method can trivially be smoothed in a post process. However, this is not always necessary. Figure~\ref{fig:smoothness} (a) shows the result of Skeletonizing a Dupin cyclide. The figure shows both the separators (color coded) and the resulting skeleton which is quite smooth due to the regularity of the cyclide and its sampling which together ensure that the obtained separators are perfectly circular. It should be noted that to obtain this result, we apply the optimization (Section~\ref{sec:optimizing-minimal-separators}) which we do not enable for the other examples, since the effect is only noticeable on a regular structure such as the cyclide.

Figure~\ref{fig:smoothness} (b) demonstrates that we can also obtain a very regular skeleton for slightly less perfect input. In this case, edges of the thin part of the cyclide have been contracted (here the separators are individual vertices) and a section of the thick part has been cut out (here the separators are open circular arcs), but the result is still a very regular skeleton. 

\begin{figure}[h!]
    \centering
    \includegraphics[width=\columnwidth]{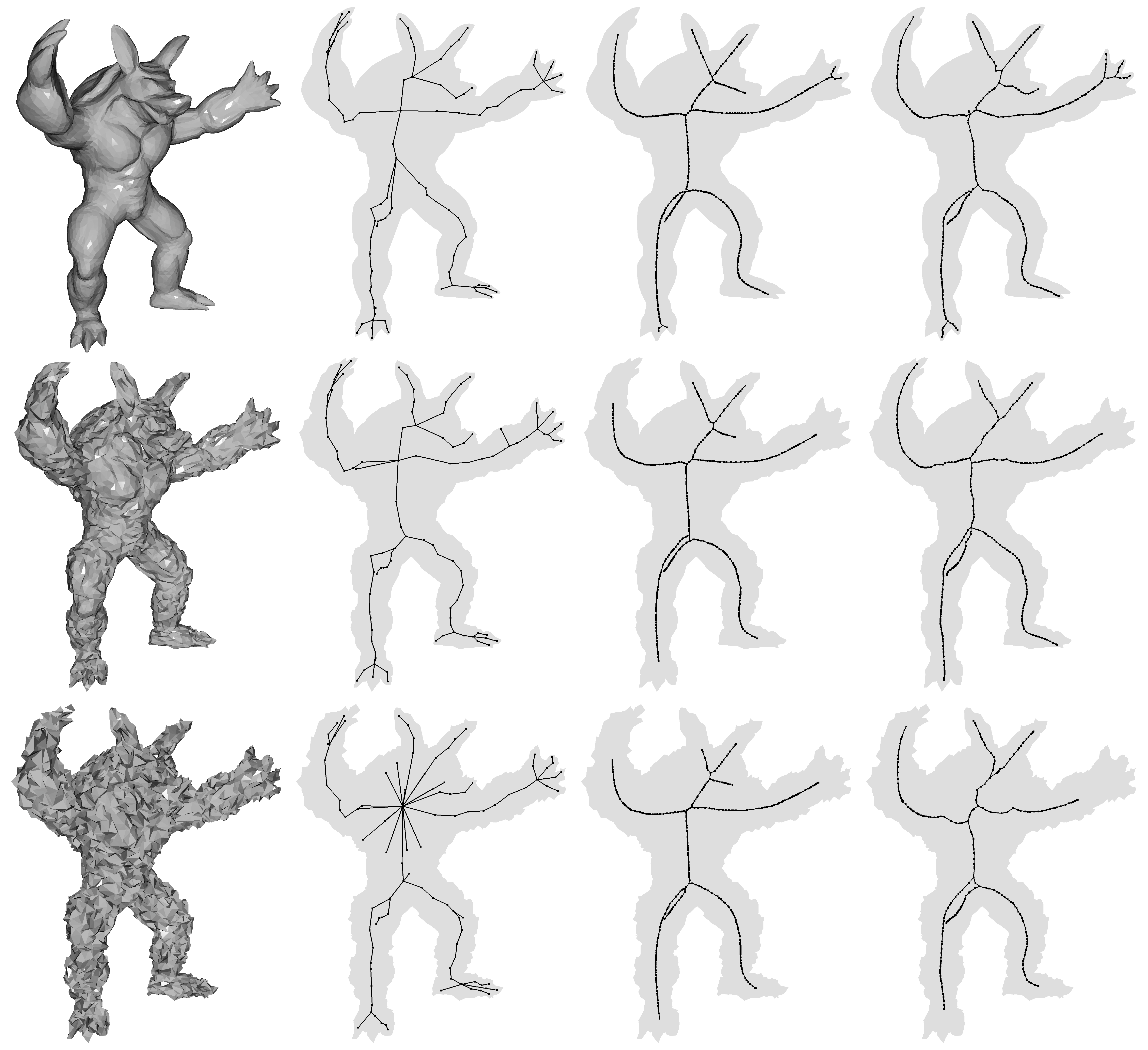}
    \caption{From left to right these images show: the input mesh, the LS skeleton, the MCF skeleton with $w_H=0.1$, and with $w_H=1$. The rows contain the Armadillo without noise (top row), the same model after one application of vertex displacement noise in the middle row, and two applications of random displacement (bottom row).}
    \label{fig:noise}
\end{figure}

Figure~\ref{fig:smoothness} (c \& d) show the Armadillo and the human hand models skeletonized with 0, 1, 2, or 3 steps of smoothing. In both cases, smooth skeletons result. We note that the effect seems to be purely beneficial for the hand, but in the case of the Armadillo, some distortion of fine features takes place, comparable to what we might see using the MCF method: cf. Netune's hand and trident in Figure~\ref{fig:mesh-comparison}. We can also compare to the Armadillo skeleton produced by MCF which is shown in Figure~\ref{fig:noise}, but since MCF fails to capture several digits, it is not possible to make a direct comparison.
\paragraph{Blending Reeb Graphs}
\begin{figure*}[h!]
    \centering
    \includegraphics[width=\textwidth]{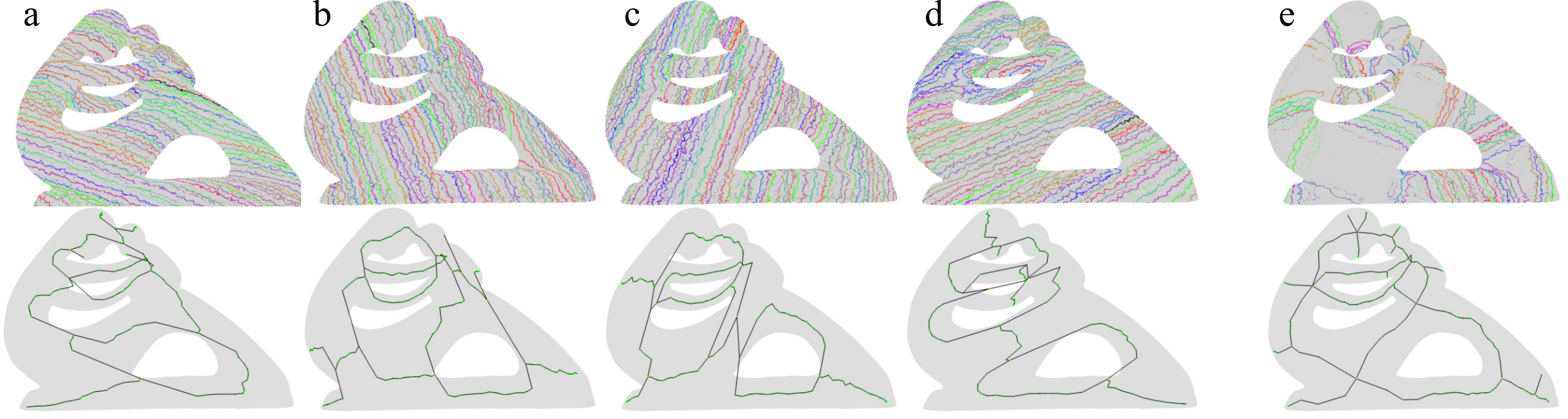}
    \caption{a,b,c, and d show separators (top) computed from height functions associated with four different directions. The resulting skeletons (below) are tantamount to Reeb graphs for these height functions. e shows the result of packing the "Reeb separators" using our separator packing approach. The result is skeleton that is less anisotropic and better captures the overall shape.}
    \label{fig:reeb-blending}
\end{figure*}
Given a shape, $S$, and a height function, $h: S \rightarrow \R$, the Reeb graph of $S$ with respect to $h$ is the quotient space of $S$ under the equivalence relation that two points $x$ and $y$ are equivalent if they belong to the same connected component of the same level set, i.e. $h(x)=h(y)$.

Given a function $h$ and a spatially embedded graph, we can easily extract a set of separators using an approach that is similar to the discrete contour computation algorithm of Tierny et al. \cite{tierny2008enhancing}. If a value of $h$ is associated with each vertex, we proceed by marking the vertices which are discrete minima as sources and then we add all their neighbors to a queue, $Q$, of vertices being visited. At each step, we ``freeze'' the vertex in $Q$ with the smallest value and add all its unvisited neighboring vertices to $Q$. This is similar to the common implementation of Dijkstra's algorithm, except that we are not computing distances but merely tracking an existing function. Importantly, at any time step, the vertices in $Q$ form a separator (between frozen vertices and unvisited vertices) whose connected components can be used as local separators. By keeping track of the time step at which vertices are added and removed, we can easily output non-overlapping local separators, and the result is tantamount to a discrete Reeb graph.

Four such separator collections (and associated skeletons) are shown in Figure~\ref{fig:reeb-blending}(a-d). The height functions employed are simply the vertex positions projected onto an axis and subsequently smoothed. 
Applying our packing algorithm, we can now compute a set of combined separators and the resulting skeleton shown in Figure~\ref{fig:reeb-blending}e. In effect, our separator packing algorithm blends the four input Reeb graphs, and the result is a skeleton that appears to better capture the overall shape than the individual Reeb graphs. In particular, centeredness (iii) is improved.

Clearly, we could also have employed a number of other functions such as eigenfunctions of the graph Laplacian or geodesic distance functions as choices of $h$. We did experiment with all of these options, individually and in combination. However, in our experience, local separators are better at capturing fine details of the geometry. 
\paragraph{Noisy Meshes}
We compared LS to MCF in terms of response to noise. The vertices of the Armadillo were corrupted by adding a random vector sampled from a ball of radius equal to half the average edge length. This procedure was also applied a second time to measure the response to more extreme noise. The results are shown in Figure~\ref{fig:noise}. We note that the effect of noise on the LS method appears to mostly be spurious leaf nodes, which could be removed fairly easily, whereas MCF seems to loose rather than gain features. This is unsurprising since MCF is based on smoothing, and more smoothing is probably needed in order to obtain a skeleton from a noisy model.  
\paragraph{Quadrilateral Meshes}
\begin{figure}[t]
    \centering
    \includegraphics[width=\columnwidth]{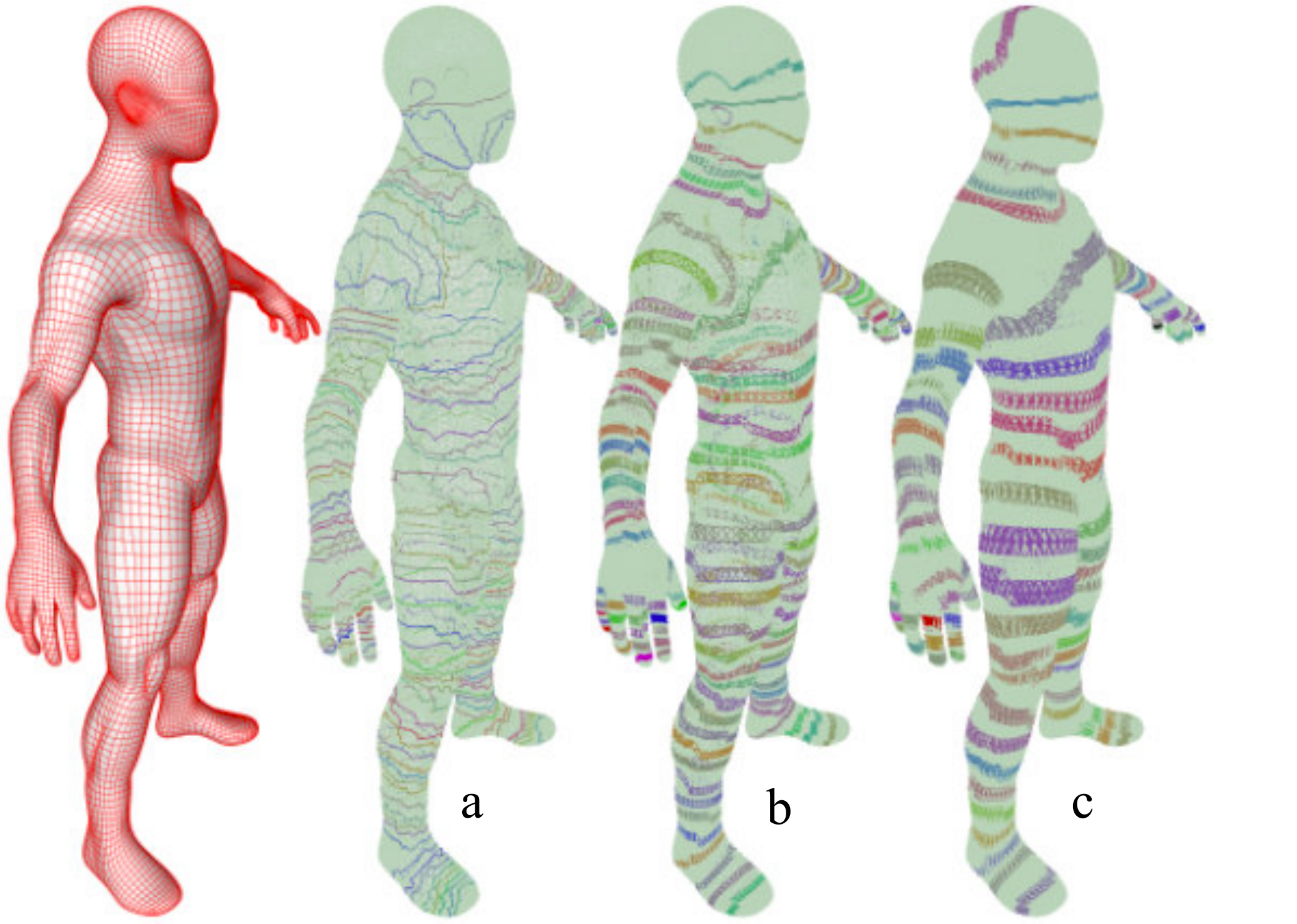}
    \caption{The effect of increasing connectivity. The original quad mesh is shown left, (a) shows the separators  for a graph constructed from a triangulation of the quad mesh, (b) shows separators for a graph constructed by connecting each vertex, $v$, to $N^2(v)$, and (c) shows the corresponding result for $N^3(v)$.}
    \label{fig:quad-saturation}
\end{figure}

If a pure quadrilateral mesh is provided as input to our method, the output is an identical quad mesh since each vertex will be a separator. Simply triangulating the quad mesh makes the algorithm behave as expected. However, we can also increase the connectivity in other ways than by triangulating. A simple procedure is to connect all vertices to the neighbors of their neighbors. More generally, we can define $N^k(v)$ as the set of vertices at a distance of at most $k$ edges. Figure~\ref{fig:quad-saturation} shows the skeletons produced from a triangulation (a) and from a graph where each vertex, $v$, has been connected to $N^2(v)$ (b) and $N^3(v)$ (c). Unsurprisingly, the separators become fewer and thicker as the connectivity increases.
\subsection{Voxel Grids}
Like a triangle mesh, a voxel grid can be construed as a graph. In this case, the positions of the graph vertices are simply the positions of the voxels in the 3D grid, but we remove vertices that correspond to background voxels and keep only those which correspond to foreground (or object). Each graph vertex should be connected to all the vertices which correspond to foreground vertices in its 26-connected neighborhood. If we think of a voxel as a small cube in a grid of cubes, 26-connectivity means that the voxel (and hence the graph vertex) is connected to all the voxels with which it shares a face, edge, or a vertex.

\begin{figure*}[t]
    \centering
    \includegraphics[width=\textwidth]{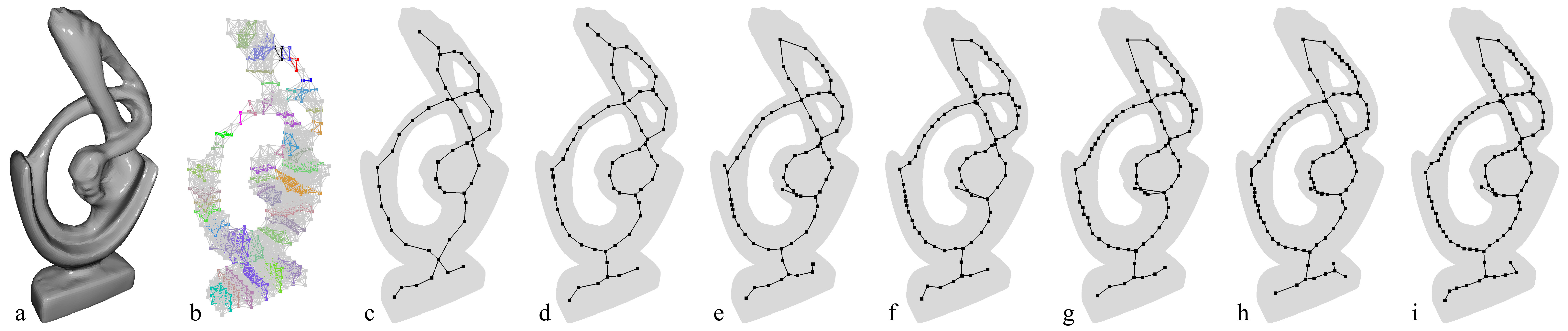}
    \caption{Volumetric skeletons generated from the Wood Sculpture. The original mesh (a), a graph containing 760 vertices created from a voxel grid with local separators shown in color (b). The skeleton produced from this graph (c), the skeleton produced from a volumetric graph consisting of 1800 vertices (d), 3540 vertices (e), 6166 vertices (f), 9830 vertices (g), 14723 vertices (h), and 
    20966 vertices (i).}
    \label{fig:wood-sculpture-volumetric}
\end{figure*}
We conducted two experiments using voxel grids. In the first experiment, we sampled the Wood Sculpture model on a voxel grid in order to produce a graph. We did this for seven voxel grids with longest side lengths ranging from 30 through 90. The results are shown in Figure~\ref{fig:wood-sculpture-volumetric}. There is a clear variation in the skeletons, and comparing with Figure~\ref{fig:mesh-comparison} it seems that slightly more details are captured in the mesh skeleton. This is to be expected since the voxel grids are derived from the mesh. Note how the local separators shown in Figure~\ref{fig:wood-sculpture-volumetric}(b) are laminar rather than cycles.

\begin{figure}[h!]
    \centering
    \includegraphics[width=\columnwidth]{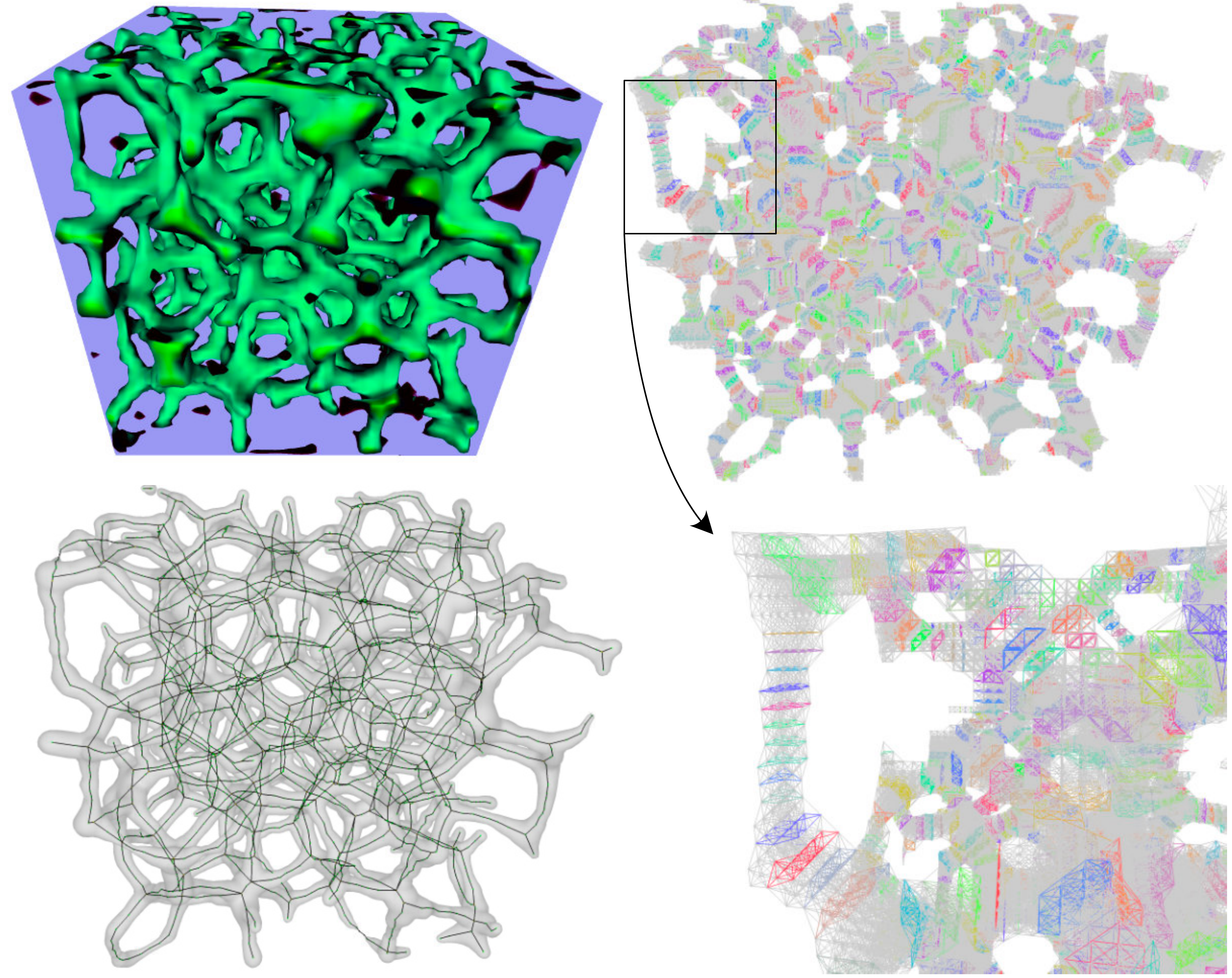}
    \includegraphics[width=\columnwidth]{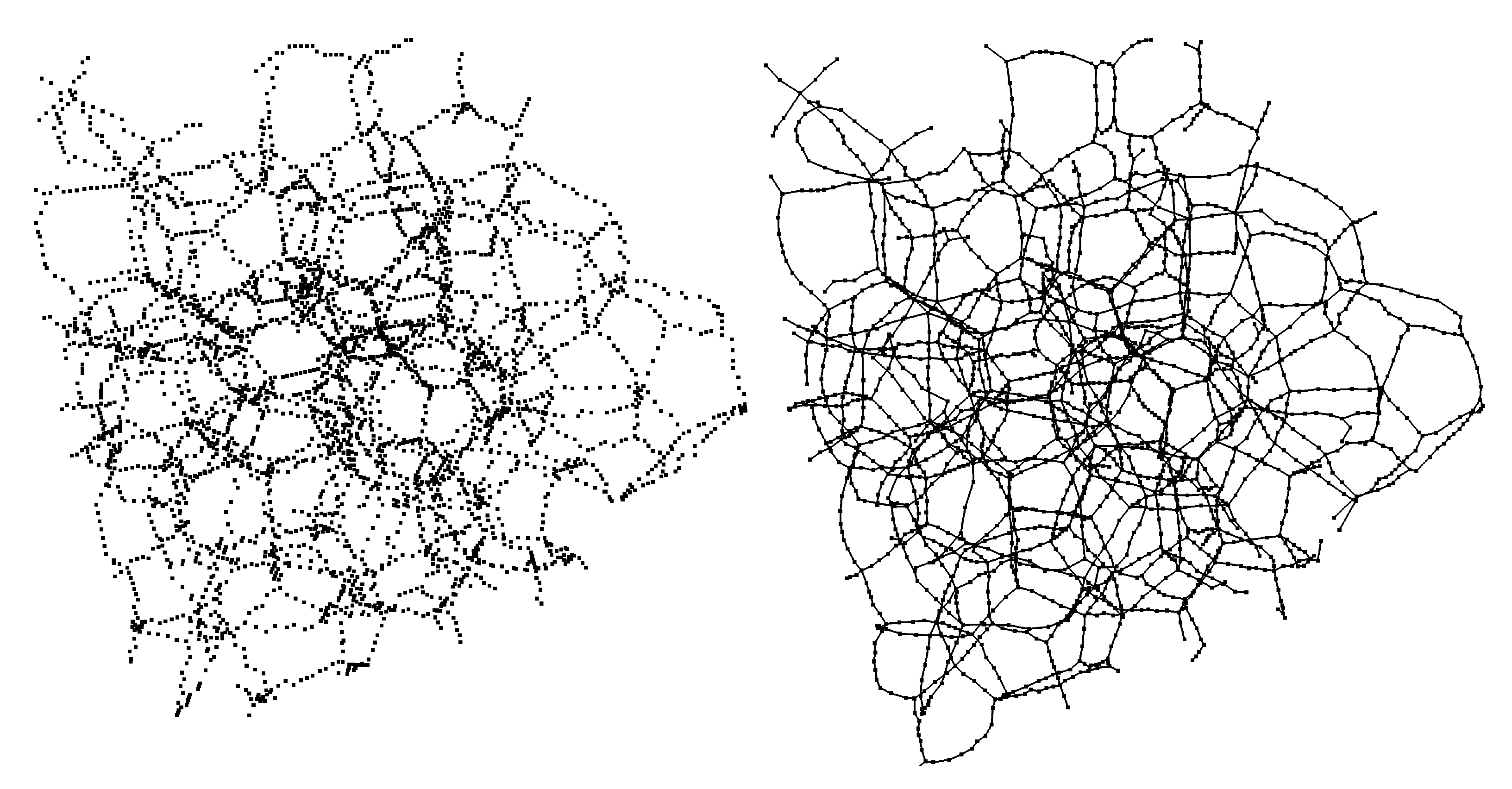} 
    \caption{Skeleton from tomographic reconstruction of foam. Clockwise from the top left image, the first two rows show a volume rendering, the local separators, a detail of the local separators, and the reconstructed skeleton. The bottom row shows a comparison between morphological thinning (left) and our method.}
    \label{fig:tomofoam}
\end{figure}

In another experiment, we ran the code on X-ray CT data of liquid foam from the TomoBank repository \cite{deCarlo2018tomobank}. This 3D image is of size $80\times 80\times 80$ and we create graph nodes for voxels above 0.37 corresponding to 11.3\% of the voxels becoming graph nodes. The output is shown in Figure~\ref{fig:tomofoam} (top two rows).
We can easily compare our approach to image based approaches to skeletonization. The scikit-image (\url{https://scikit-image.org}) package for Python contains a function which, given a 3D binary image, returns a new 3D binary image containing a 1 voxel thick skeleton. This function works by iterative thinning according to a method by \cite{lee1994building}, and it is extremely fast, producing a skeleton in roughly 55 milliseconds. However, the output is simply a binary image whereas our method estimates more precise vertex positions and their connectivity. A comparison is shown in Figure~\ref{fig:tomofoam} (bottom row).
%
%
\subsection{Point Clouds}
In some respects, point clouds are a more challenging type of data than meshes or voxel grids. Since our algorithm operates on graphs, we need to convert an input point cloud to a graph before proceeding. Figure~\ref{fig:cartoon} shows a simple example of a 2D point cloud where a graph has been constructed by connecting each point to all neighbors in a given radius, and then our method has been applied to the graph. 

Of course, Figure~\ref{fig:cartoon} is a toy example. For real point clouds, we use the following pipeline:
\begin{enumerate}
    \item Construct a graph, $G$ by connecting each point to its at most k-nearest neighbors within a given radius.
    \item Connect each vertex, $v \in G$ to vertices in $N^l(v)$ for some $l<10$ (again within a certain radius).
    \item Simplify the graph by edge contraction.
\end{enumerate}
We will briefly justify this pipeline. In order to separate features of the represented object (say two branches of a tree) we often need to use a radius in Step 1 that is too small to ensure that we obtain all edges needed to fulfill the reconstruction condition. However, as long as these edges can be reached by going from neighbor to neighbor, we can obtain them via the second step. Thus, it is possible to maintain a gap between features smaller than the greatest distance between vertices that should be connected. Often the resulting graph contains more vertices than needed to capture the skeletal features we seek, hence the third step. Note that while this pipeline is not trivial and depends on several parameters that must be determined, it is simpler than any method that are aware of for reconstructing a manifold triangle mesh for an object of unknown genus.

A LiDAR scan of a botanical tree consisting of 832943 points was provided by collaborators. Applying the pipeline above, we obtained a graph consisting of 48661 vertices. Once the LS method had been applied to this graph, the skeleton was garbed using convolution surfaces \cite{bloomenthal1991convolution} producing the result shown in Figure~\ref{fig:tree-lineup}.

\begin{figure*}
\includegraphics[width=\textwidth]{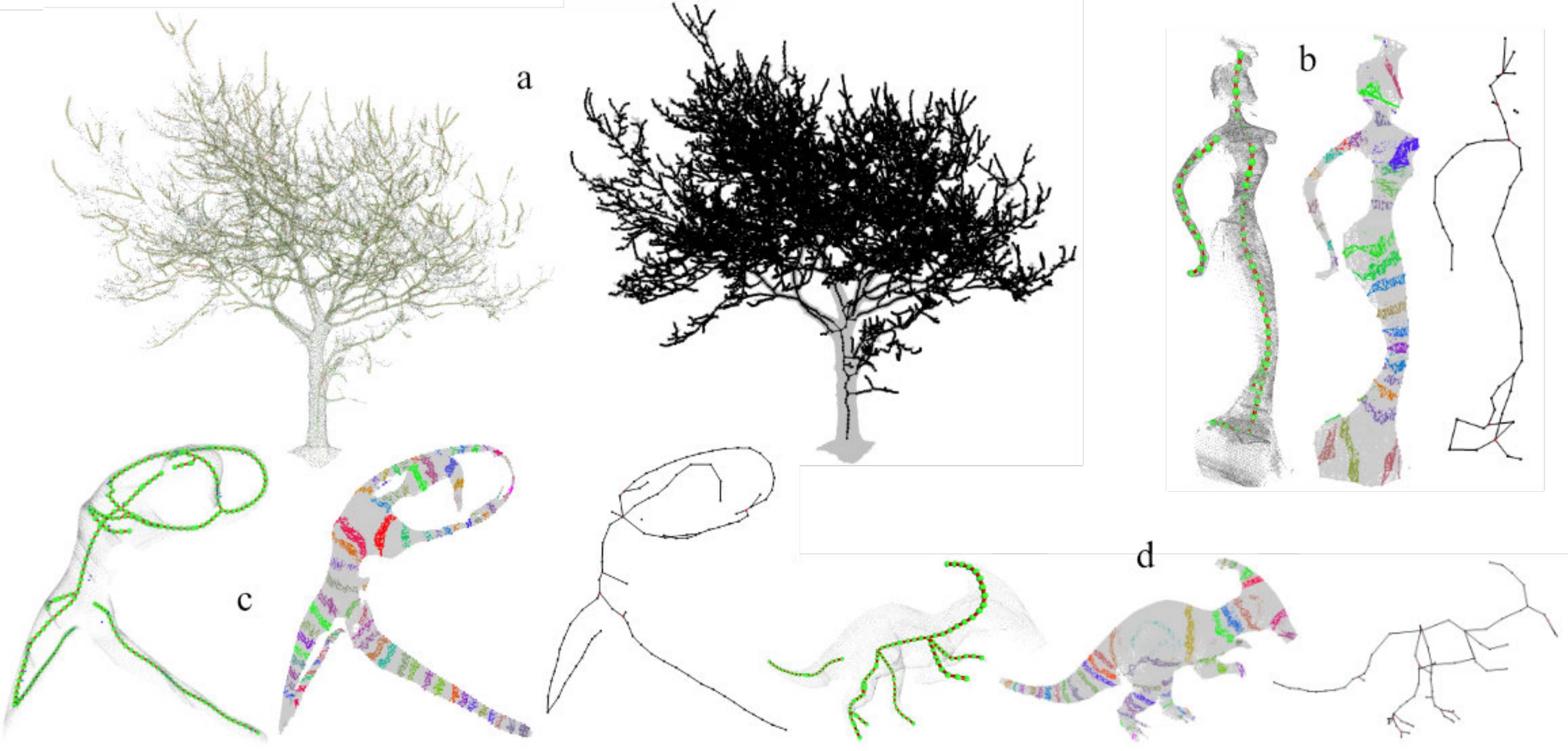}
\caption{(a) shows a comparison using the botanical tree point cloud. On the left, the L1M algorithm by Huang et al. was employed \cite{huang2013l1}. On the right, our local separator algorithm was used to skeletonize the graph produced from the points. (b,c,d) show a comparison between the method of Huang et al. and our method for three point clouds. In each case, the left image was computed using L1M, the middle image shows the input graph and the color coded separators found using the LS method, and the right image shows the final LS skeleton.}
\label{fig:tree}. 
\end{figure*}

Using the botanical tree and three of their provided example point clouds as test data, we compared our LS method to the L1-medial skeleton method (L1M) by Huang et al. \cite{huang2013l1}. The comparison was carried out using the executable provided by the authors. Table~\ref{tab:L1M} summarizes the comparison. Note that we do not compare timings since the tests were carried out on different computers using different numbers of vertices and samples respectively, for LS and L1M. The results are shown in Figure~\ref{fig:tree}. It is clear that the two methods result in skeletons which are qualitatively very different. L1M is not constrained by point connectivity. This means that it can sometimes capture a feature defined by several disjoint point clusters with a single chain of skeletal edges. This is evident in Figure~\ref{fig:tree} (d). Due to several gaps in the point cloud, the back and the belly of the Dinosaur are represented by different edge chains using our method, but captured with a single chain of edges using L1M. However, L1M does have a propensity for creating gaps in the skeleton, and this is very apparent in the case of the botanical tree where the method seems to invent some branches while missing other. Admittedly, L1M is governed by numerous parameters. We tried eight different settings on the tree model but were unable to obtain a better result.
\begin{table}
\begin{tabular}{l|r|r|rr}
&  & \multicolumn{1}{c}{LS} & \multicolumn{2}{|c}{L1M} \\
Model & Points & vertices &  samples &  iter \\
\hline
Tree       & 48661& 48661 & 48661 & 500\\
Dinosaur &  25713 & 2329 & 25713 & 255\\
GLady     &  44230 &  859 & 5559 & 165\\
Yoga3 	&  21903 & 2307 & 21903 & 199\\        
\end{tabular}
\caption{Comparison of the LS and L1M methods for skeletonization. This table contains input sizes and the number of iterations used by L1M.} 
\label{tab:L1M}           
\end{table}

\subsection{Experimental running time analysis.}

To experimentally evaluate the performance of our algorithm, we ran it on subsamples of \emph{Wood Sculpture} (see Figure~\ref{fig:mesh-comparison}). We ran two experiments: For one, we thinned out the mesh grid to obtain shapes of fewer vertices (all, half, quarter, etc, down to $(1/2)^{-6}$). Secondly, we transformed the mesh to a voxel grid with a variable number of voxels. 

In all experiments, we found that the time spent computing the separators by far dominates the running time (>95\% of the time was spent on finding separators as opposed to packing and extraction), and that the proportion of total time spent on computing separators tends to 1 as the number of vertices grows.

\begin{figure}[h]
    \includegraphics[width=\columnwidth]{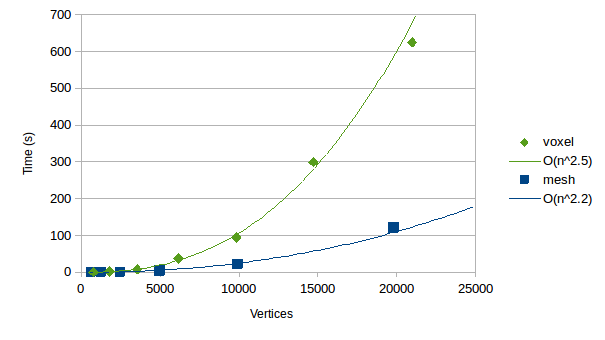}
\caption{The running time for submeshes formed from Wood Sculpture, and their power regression curves. 
\label{fig:totaltime}}
\end{figure}

Although the running time is indeed a function of two parameters, the number of vertices $n$ and the average separator size $\Sigma_{\max}$ found in Algorithm~\ref{alg:local-separator}, we run the experiment only on this one figure (of fixed lankiness), and analyse the running time simply as a function of $n$. We expected a running time of at most $O(n^{2.5})$ for the mesh, and $O(n^{2+\frac{2}{3}})$ for the voxels, though hopefully less because of the sampling. As shown in Figure~\ref{fig:totaltime}, we did indeed observe an asymptotically improved running time compared to the worst-case, and we did  observe that voxels were asymptotically slower than meshes.

A natural question, since the running time depends on the number of vertices sampled, is: how many vertices are sampled? Here, we were expecting asymptotically fewer points to be sampled from the voxel grid in comparison to the mesh. Indeed, this appears to be the case
. Using power regression, we estimate a total sample of $O(n^{.74})$ for an $n$-vertex mesh, and $O(n^{.66})$ for an $n$-vertex voxel grid. Note, however, that these numbers are highly dependent on the lankiness of the specific shape in question.




%% file: Discussion.tex
\section{Conclusions and Future Work}
We have presented our results for a skeletonization algorithm based on the notion of local separators. Our method accepts as input a graph whose vertices map to spatial positions. While certain assumptions on connectivity must be met for a meaningful output to be produced, our method works on a wide range of inputs, and the resulting skeletons tend to resolve all important features.

A local separator can be seen as a skeletal atom, and a particular point of our method is that we find these atoms first and then assemble the skeleton. This is somewhat different from most methods based e.g. on clustering~\cite{jiang2013curve}, contraction~\cite{huang2013l1} or smoothing~\cite{Tagliasacchi12Mean,au2008skeleton} where these two steps are coupled, meaning that we only know the atoms when we have found the skeleton (at least locally). In our approach, we require the atoms to be separators and find a super set of the needed atoms, i.e minimal local separators. We speculate that this is what allows the LS approach to resolve finer details in the shapes we tested than MCF~\cite{Tagliasacchi12Mean}: if the skeletal atoms are a product of a process which terminates only when the skeletal structure has been computed, then atoms corresponding to small structures might have been incidentally removed when that point is reached. Of course, the advantage is only reaped because the LS algorithm is able to resolve small details in the first place. This is thanks in large part to the notion of separator minimality.

Unlike a number of approaches to curve skeletonization, notably \cite{dey2006defining}, our method does not take the medial surface into account. However, recent methods compute curve skeletons which are attracted to the medial surface, and this would certainly be possible also with our method and might be worthwhile investigating. Likewise, smoothing is not inherent in our method. Arguably, this is an advantage since smoothness can also make the skeleton less precise, but both attraction to the medial surface and optional smoothing could be combined leading to a method that would flexibly allow the user to obtain skeletons with desired degrees of smoothness and proximity to the medial surface in the vein of \cite{Tagliasacchi12Mean} - at least for shapes where the medial surface is well defined.

%

In this work, we dealt with performance through parallelization and sampling, but the process of finding and packing local separators becomes very slow for large separators. 
The use of a dynamic spatial data structure~\cite{Arya1998approximate} when searching the front, and a structure for dynamic connectivity of the front~\cite{HolmLT2001}, would reduce the search time in Algorithm~\ref{alg:local-separator} from linear to polylogarithmic in the number of vertices. 

The performance of the sampling scheme could also be improved. In particular, we might not have to start from a vertex when computing every single separator. In fact, given an existing separator, the connected components of its neighbors are also separators in most cases. Thus, it is, in fact, quite possible to find separators by a search that starts from a single separator and propagates out in this way. However, our preliminary experiments indicate that sampling is still required, and it is not easier to find small features using this approach. Thus, more work is required to find a good sampling scheme that balances coverage with the use of the information in the separators already found.

An important feature of our algorithm is that it works for any input that is a graph or that can easily be transformed into a graph. As such, it may be used for skeletonization of shapes not only in $\mathbb{R}^2$ and $\mathbb{R}^3$, but shapes in higher dimensions, or non-euclidean graphs. 
An open question for future work is to find such applications and test our algorithm on those.